\documentclass[prb, superscriptaddress, showpacs, twocolumn]{revtex4}  
\usepackage{amssymb}
\usepackage{amsmath}
\usepackage{bbm}
\usepackage{graphicx}
\usepackage{color}

\begin{document}

\author{Philipp Werner}
\affiliation{Department of Physics, University of Fribourg, 1700 Fribourg, Switzerland}
\author{Hugo Strand}
\affiliation{Department of Quantum Matter Physics, University of Geneva, 24 Quai Ernest-Ansermet, 1211 Geneva 4, Switzerland}
\author{Shintaro Hoshino}
\affiliation{Department of Physics, Saitama University, Saitama 338-8570, Japan}
\author{Yuta Murakami}
\affiliation{Department of Physics, University of Fribourg, 1700 Fribourg, Switzerland}
\author{Martin Eckstein}
\affiliation{Department of Physics, University of Erlangen-N\"urnberg, 91058 Erlangen, Germany}

\title{Enhanced pairing susceptibility in a photo-doped two-orbital Hubbard model}

\date{\today}

\hyphenation{}

\begin{abstract} 
Local spin fluctuations provide the glue for orbital-singlet spin-triplet pairing in the doped Mott insulating regime of multi-orbital Hubbard models. At large Hubbard repulsion $U$, the pairing susceptibility is nevertheless very low, because the pairing interaction cannot overcome the suppression of charge fluctuations. Using nonequilibrium dynamical mean field simulations of the two-orbital Hubbard model, we show that out of equilibrium the pairing susceptibility in this large-$U$ regime can be strongly enhanced by creating a photo-induced population of the relevant charge states, and that this susceptibility correlates with the local spin susceptibility. Since a strong enhancement of the pairing requires a low kinetic energy of the charge carriers, the phenomenon is supported by the ultra-fast cooling of the photo-doped carriers through the creation of local spin excitations.
\end{abstract}

\pacs{ 71.10.Fd} 

\maketitle

\section{Introduction}

Inducing or stabilizing  electronic orders 
by driving a correlated electron system into a nonthermal state is a new and promising strategy for the control and manipulation of material properties. Examples include the observation of light-induced superconductivity in phonon-driven cuprates \cite{Fausti2011,Kaiser2014} and fulleride compounds,\cite{Mitrano2016} as well as the enhancement of excitonic order by photo-excitation of electron-hole 
pairs.\cite{Mor2017} 
While several theoretical proposals for the enhancement of excitonic and superconducting condensates have been put forward,\cite{Murakami2017ei,Sentef2016,Okamoto2016,Kim2016,Mazza2017,Ido2017,Knap2016,Komnik2016,Babadi2017,Kennes2017,Sentef2017,Nava2017} most 
scenarios can be understood within an equilibrium picture, because they rely on changes of the bandwidth\cite{Sentef2016,Ido2017} or interaction parameters,\cite{Kim2016,Mazza2017} or they do not fully take into account the competing effects of heating and nonthermal energy distributions  
 in driven systems.\cite{Murakami2017ei,Kim2016,Knap2016,Komnik2016,Babadi2017,Kennes2017,Sentef2017,Nava2017}  
Since heating effects can strongly reduce electronic orders in driven systems,\cite{Murakami2017floquet} 
interesting transient states can be expected to occur in situations where the energy absorption is minimal (e.g. in the case of sub-gap driving \cite{Fausti2011}), 
or where a large part of the injected energy is transiently stored as potential energy.\cite{Rosch2008,Werner2012} Here, we demonstrate an example of the second type, namely a strongly enhanced spin-triplet pairing susceptibility in a photo-doped two-orbital Hubbard model with a large Mott gap. This enhancement can neither be explained by chemical doping of the Mott insulating parent state, nor by increasing temperature, and hence is a genuine nonequilibrium effect. 

Multi-orbital Hubbard models with Hund coupling exhibit an orbital-singlet spin-triplet superconducting phase at low temperature.\cite{Klejnberg1999,Zegrodnik2013,Koga2015,Hoshino2015} In equilibrium and at intermediate Hubbard interaction, this pairing instability is closely connected to the spin-freezing crossover\cite{Werner2008} that occurs as the half-filled Mott insulator is approached by changing the filling at fixed interaction strength.\cite{Hoshino2015,Hoshino2016} In the spin-freezing crossover regime, slowly fluctuating local moments appear, which induce pairing at low temperature. Closer to half-filling, the local moments freeze and the resulting incoherent metal state is characterized by a suppressed pairing susceptibility. 
In the two-orbital model, at large enough Hubbard repulsion, a Mott insulating phase is realized also at 3/4 filling ($n=3$) and 1/4 filling ($n=1$). While the local spin susceptibility shows an enhancement 
near $n\approx 1.5$ and $n\approx 2.5$, 
the pairing susceptibility  
remains very low in 
the whole filling range $1<n<3$. 
This indicates that in this region of the phase diagram, the pairing interaction cannot overcome the suppression of charge fluctuations by the interaction $U$.

As we will show in this study, the injection of triplon and singlon charge carriers into the half-filled Mott insulator in this strongly correlated regime creates a photo-doped metal state with a strongly enhanced pairing susceptibility. 
The increased density of singlons and triplons enables the charge fluctuations associated with pairing in a half-filled system, while local spin-flip excitations cool the photo-doped carriers \cite{Strand2017} down to a temperature of the order of the Hund coupling within a few inverse hopping times. Further cooling by a boson bath results in a substantial increase of the pairing susceptibility with respect to the values obtainable by chemical doping in equilibrium.
The phenomenon is related to the metastable superfluidity in a strongly repulsive Hubbard model discussed in Ref.~\onlinecite{Rosch2008}, but it is demonstrated here for a moderate density of photo-doped carriers that move in a background of half-filled doublon states, and with a formalism that captures the heating effect associated with the photo-doping process.

The paper is organized as follows: In Sec.~\ref{sec:method} we describe the model and the method used to calculate the nonequilibrium spin-triplet pairing susceptibility 
and the local 
spin susceptibility. Section~\ref{sec:equilibrium} investigates the relation between pairing and spin susceptibility in the equilibrium model, while Sec.~\ref{sec:nonequilibrium} discusses the effect of photo-doping on these susceptibilities in the strongly interacting Mott regime. A summary and conclusions are presented in Sec.~\ref{sec:summary}. Appendix~\ref{appendix} contains a derivation of the dynamical mean field self-consistency equation for the two-orbital Hubbard model with spin-triplet superconductivity.

\section{Model and Method}
\label{sec:method}

We consider a two-orbital Hubbard model with Hamiltonian 
\begin{align}
&H_\text{latt}(t) = \sum_{i \ne j} \sum_{\alpha=1,2} \sum_\sigma v^\alpha_{ij}(t)  c^\dagger_{i,\alpha\sigma} c_{j,\alpha\sigma}\nonumber\\
&\hspace{6mm}+\sum_i \sum_{\alpha=1,2}\Big[U n_{i,\alpha\uparrow} n_{i,\alpha\downarrow}-\mu (n_{i,\alpha\uparrow}+n_{i,\alpha\downarrow})\nonumber\\
&\hspace{25mm}+B_z(t) (n_{i,\alpha\uparrow}-n_{i,\alpha\downarrow})\Big] \nonumber\\
&\hspace{6mm}+\sum_i \sum_\sigma \Big[(U-2J) n_{i,1\sigma} n_{i,2\bar\sigma} + (U-3J) n_{i,1\sigma} n_{i,2\sigma}\Big] \nonumber\\
&\hspace{6mm}+H_\text{pair-field},
\label{H}
\end{align}
where $v^\alpha_{ij}$ is the hopping amplitude between sites $i$ and $j$ for orbital $\alpha$, $\sigma$ denotes spin, $\mu$ the chemical potential, $B_z$ the magnetic field, $U$ the intra-orbital interaction, and $J$ the Hund coupling. 
Spin-flip and pair-hopping terms are neglected in this study. 
(Pair-hopping is irrelevant for $J>0$.  Spin-flips would typically suppress spin-triplet superconductivity,\cite{Hoshino2015} but neglecting them can be viewed as the introduction of a spin anisotropy which may originate from spin-orbit coupling.)  
To measure the spin-triplet pairing susceptibility, we also add  
a pair-field term
\begin{equation}
H_\text{pair-field}=-\sum_i P(t)(c^\dagger_{i,1\uparrow}c^\dagger_{i,2\uparrow}-c^\dagger_{i,1\downarrow}c^\dagger_{i,2\downarrow} + h.c.).
\end{equation}
For the lattice we assume a Bethe lattice 
with infinite coordination number. 

To investigate the nonequilibrium properties of this model, we use the nonequilibrium extension of dynamical mean field theory.\cite{Freericks2006,Aoki2014} In dynamical mean field theory (DMFT),\cite{Georges1996} the lattice model (\ref{H}) is mapped to a 2-orbital quantum impurity model with action
\begin{align}
S=& 
\int_\mathcal{C} dt dt' 
\psi^\dagger(t)\Delta(t,t')\psi(t')
+ \int_\mathcal{C} dt H_\text{loc}(t),
\end{align}
where the local terms represented by $H_\text{loc}$ are identical to those of the lattice model (\ref{H}), and in the hybridization term, we introduced the four-component spinor $\psi^\dagger=(c_{1\uparrow}^\dagger, c_{2\uparrow}, c_{1\downarrow}^\dagger, c_{2\downarrow})$ as well as the matrix valued hybridization function 
\begin{equation}
\Delta(t,t') = \left(
\begin{array}{cccc}
\Delta^{c^\dagger  c}_{1\uparrow 1\uparrow} & \Delta^{c^\dagger c^\dagger }_{1\uparrow 2\uparrow} & 0	& \Delta^{c^\dagger c^\dagger }_{1\uparrow 2\downarrow} \\
\Delta^{cc}_{2\uparrow 1\uparrow} & \Delta^{c c^\dagger }_{2\uparrow 2\uparrow} & \Delta^{cc}_{2\uparrow 1\downarrow} & 0 \\
0	& \Delta^{c^\dagger c^\dagger }_{1\downarrow 2\uparrow} & \Delta^{c^\dagger  c}_{1\downarrow 1\downarrow} & \Delta^{c^\dagger c^\dagger }_{1\downarrow 2\downarrow} \\
\Delta^{cc}_{2\downarrow 1\uparrow} & 0 & \Delta^{cc}_{2\downarrow 1\downarrow} & \Delta^{c c^\dagger }_{2\downarrow 2\downarrow}
\end{array}
\right).
\end{equation} 
(Note that we allow for anomalous components associated with inter-orbital pairing, but set those associated with intra-orbital pairing to zero.)
The hybridization function $\Delta$ 
is determined self-consistently in such a way that the impurity model Green's function  
\begin{align}
G(t,t')&=-i\langle \mathcal{T}\psi(t) \psi^\dagger(t')\rangle_S \nonumber\\
&= \left(
\begin{array}{cccc}
G^{c c^\dagger}_{1\uparrow 1\uparrow} & G^{cc}_{1\uparrow 2\uparrow} & 0	& G^{cc}_{1\uparrow 2\downarrow} \\
G^{c^\dagger c^\dagger}_{2\uparrow 1\uparrow} & G^{c^\dagger c}_{2\uparrow 2\uparrow} & G^{c^\dagger c^\dagger}_{2\uparrow 1\downarrow} & 0 \\
0	& G^{cc}_{1\downarrow 2\uparrow} & G^{c c^\dagger}_{1\downarrow 1\downarrow} & G^{cc}_{1\downarrow 2\downarrow} \\
G^{c^\dagger c^\dagger}_{2\downarrow 1\uparrow} & 0 & G^{c^\dagger c^\dagger }_{2\downarrow 1\downarrow} & G^{c^\dagger  c}_{2\downarrow 2\downarrow}
\end{array}
\right),
\end{align}
with components $G^{ab}_{\alpha\sigma\beta\sigma'}\equiv -i\langle \mathcal{T} a_{\alpha\sigma}(t)b_{\beta\sigma'}(t')\rangle_{S}$ ($a$, $b$ stands for $c$ or $c^\dagger$), 
becomes equal to the local lattice Green's function. In the nonequilibrium version of DMFT, this self-consistent solution is computed on the Kadanoff-Baym contour $\mathcal{C}$,\cite{Freericks2006,Aoki2014} which runs from time $0$ to time $t$ along the real-time axis, back to time $0$ along the real-time axis, and then to $-i\beta$ along the imaginary time axis ($\beta$ is the inverse temperature of the system, and $\mathcal T$ the contour ordering operator). 

In a Bethe lattice with nearest neighbor hopping and coordination number $z$, the components of the $\Delta$ matrix are determined by the condition
\begin{equation}
\Delta(t,t') = \sum_{j=1}^z V_j^*(t) G_{jj}^{(0)}(t,t') V_j(t'),
\label{self_main}
\end{equation}
with $V_j(t)=\text{diag}(v^1_{j0}(t), -v^2_{0j}(t), v^1_{j0}(t), -v^2_{0j}(t))$. Here, $G_{jj}^{(0)}(t,t')$ is the cavity Green's function of the lattice, on a site next to the cavity at site 0.\cite{Georges1996} In the spirit of DMFT, we can replace $G_{jj}^{(0)}(t,t')$ by $G(t,t')$. If the hopping amplitudes are real and independent of the bond, we can furthermore replace them in the limit $z\rightarrow \infty$ by $v^\alpha_{j0}=v^\alpha_{0j}\rightarrow v^\alpha/\sqrt{z}$ to obtain the usual self-consistency condition for the infinite dimensional Bethe lattice:
\begin{equation}
\Delta(t,t') = V(t) G(t,t') V(t'),
\label{self_infinite}
\end{equation}
with $V(t)=\text{diag}(v^1(t), -v^2(t), v^1(t), -v^2(t))$. For a derivation based on the cumulant expansion, see Appendix~\ref{appendix}.

To solve the impurity problem, we use the lowest-order self-consistent hybridization expansion (non-crossing approximation, NCA).\cite{Keiter1971,Eckstein2010} This approximation is expected to be good in the Mott insulating phases, while the metallic solutions are more strongly affected by the constraint of non-crossing hybridization lines. Nevertheless, we will show that the NCA captures the qualitative features of the doping-dependent phase diagram, including the correlation between spin-triplet pairing and enhanced local spin fluctuations at moderate $U$. Most of the following analysis will be performed in the symmetric phase at $T>T_c$, and will be based on the measurement of spin-triplet pairing and local spin susceptibilities. To measure the pairing susceptibility $\chi_P$, we apply a small static pairing field $P(t)=p$ (also on the Matsubara axis) and measure the resulting order parameter 
\begin{equation}
O_P(t)=\text{Re}\langle c^\dagger_{1\uparrow} c^\dagger_{2\uparrow} \rangle=\tfrac{1}{4}\langle c^\dagger_{1\uparrow} c^\dagger_{2\uparrow}-c^\dagger_{1\downarrow} c^\dagger_{2\downarrow}+h.c.\rangle(t),
\end{equation}
which yields
\begin{equation}
\chi_P(t)=O_P(t)/p.
\end{equation}  

The dynamical contribution to the local spin susceptibility is defined in equilibrium as\cite{Hoshino2015} 
\begin{equation}
\Delta\chi=-\int_0^\beta d\tau [C_{SS}(\tau)-C_{SS}(\beta/2)],
\label{eq_delta_chi}
\end{equation}
where $C_{SS}(\tau) = -\langle \mathcal{T}_\tau S_z(\tau) S_z(0)\rangle_S$ is the local spin correlation function measured on the imaginary-time axis. On the real-time axis, and in terms of the retarded correlation function $C_{SS}^R(t) = -i\theta(t)\langle[S_z(t),S_z(0)]\rangle$, we can express the same quantity as 
\begin{equation}
\Delta\chi = \int_0^\infty dt C_{SS}^R(t)[-1+\tanh(\pi t/\beta)].
\label{Delta_chi_realtime}
\end{equation}

While the local spin correlation function can be measured directly in the NCA as a sum of bubbles of pseudo-particle Green's functions, it turns out that this approximation is very poor and not consistent with the NCA dynamics. This is evident for example by the fact that the peak of the resulting equilibrium $\Delta\chi^\text{bubble}$ does not occur in the doping region where the pairing susceptibility is largest, which is in contrast to quantum Monte Carlo (QMC) based results for moderate $U$.\cite{Hoshino2015,Hoshino2016} 
We can understand the problem by considering the case of a half-filled Mott insulator, where the dominant local states are half-filled high-spin states. The bubble approximation to the spin correlation function lacks the important contributions from time sequences  
where the spin state flips from up to down (or vice versa) 
between the two measurement times. 

To get a result which is consistent with the DMFT time evolution we apply a magnetic field pulse. 
The local susceptibility $\delta \langle S_{z,j}\rangle/\delta B_{z,j}$
on a given site $j$ can be obtained by calculating the 
time evolution of the magnetization $m(t)=\langle S_z\rangle_{S|_\Delta}(t)$ in the impurity model after a short magnetic field pulse $B_z(t)=b\delta(t-t_p)$, where  the hybridization function $\Delta(t,t')$ is fixed to the value without the magnetic field pulse. Keeping $\Delta$ fixed is consistent with the cavity construction on the infinite coordination Bethe lattice, in which a single site has a negligible effect on the hybridization function.\cite{footnote_spinsusc} 
Since the induced magnetic moment is proportional to the retarded correlation function, by multiplying it with $[-1+\tanh(\pi (t-t_p)/\beta)] $ and integrating over $t>t_p$, we obtain the following more accurate estimate of the dynamical contribution to the local spin susceptibility:
\begin{equation}
\Delta \chi^\text{pulse}=\frac{1}{b}\int_{t_p}^\infty \! dt \hspace{1mm}m(t)[-1+\tanh(\pi (t-t_p)/\beta)].
\label{eq_pulse}
\end{equation}
In practice, we integrate up to some time $t_\text{max}$ 
and apply a short magnetic field pulse of finite width, centered at $t=t_p$ and with integral $b$. 

In some calculations, we also introduce energy dissipation to a bosonic heat bath by adding a Migdal type self-energy diagram\cite{Eckstein2013} 
\begin{equation}
\Sigma_\text{boson}(t,t')_{ab}=ig^2 D_0^{(\omega_0,\beta)}(t,t')G(t,t')_{aa}\delta_{ab}
\label{eq_bath}
\end{equation}
to the diagonal components of the hybridization matrix ($a,b$ are Nambu-spin indices).
 Here, $g$ is the coupling strength and $D_0^{(\omega_0,\beta)}$ is the equilibrium free propagator for Holstein phonons with energy $\omega_0$ at inverse temperature $\beta$.\cite{Eckstein2013}  

In the rest of the paper, we consider an infinite-dimensional Bethe lattice with degenerate bands of bandwidth $W=4v$ and set $v=1$ as the unit of energy ($1/v$ as the unit of time). In the nonequilibrium simulations, the initially Mott insulating system is driven out of equilibrium by a hopping modulation $v^\alpha(t)=[1+af(t-t_p)\sin(\Omega (t-t_p))]v$ with $a$ the amplitude, $\Omega$ the driving frequency, and $f(t-t_p)$ a window function centered at $t_p$. Unless otherwise stated, we use the driving frequency $\Omega=U$, which results in an efficient photo-doping of the Mott insulator.

\begin{figure*}[t]
\begin{center}
\includegraphics[angle=-90, width=0.325\textwidth]{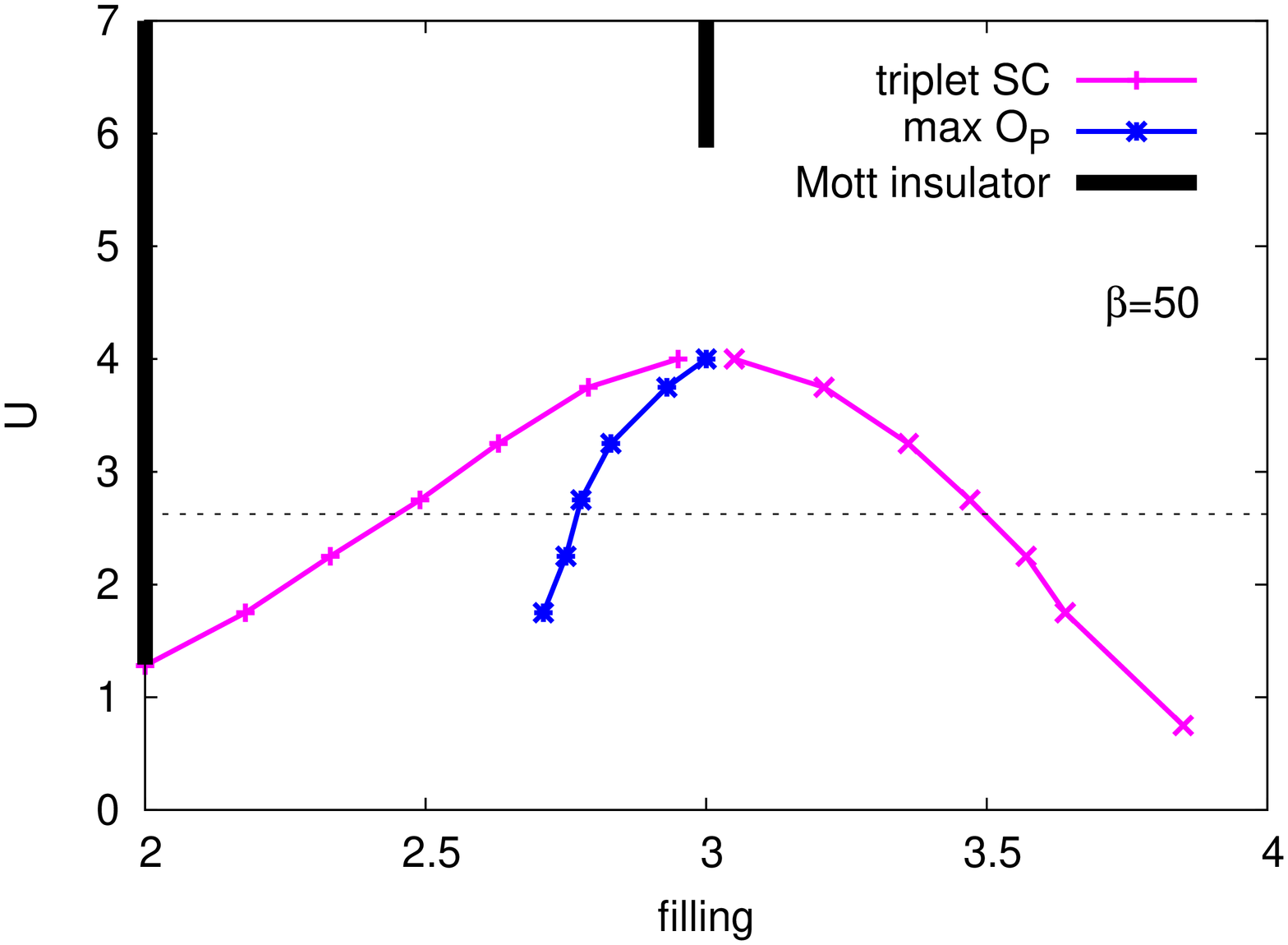}\hfill
\includegraphics[angle=-90, width=0.325\textwidth]{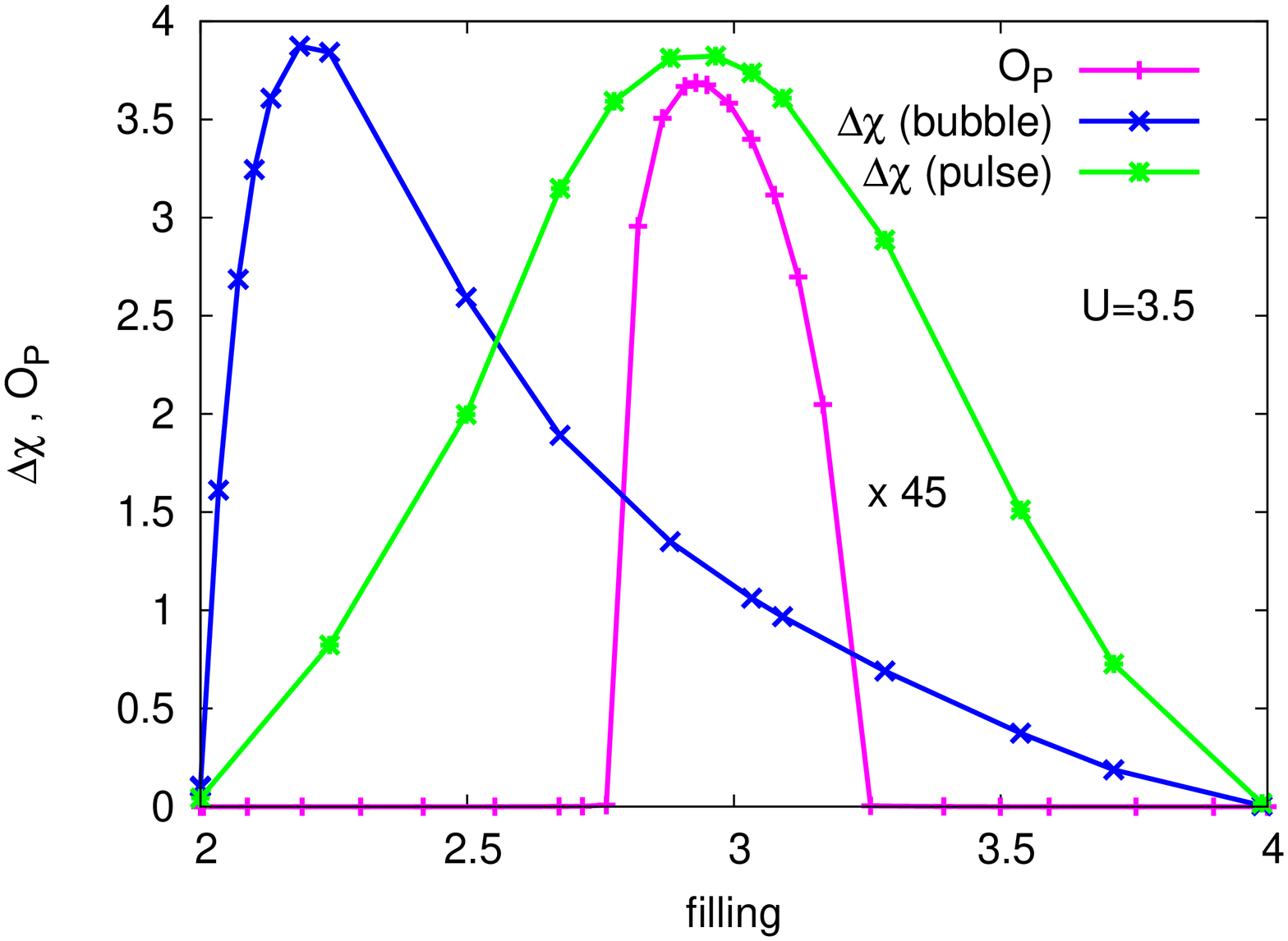}\hfill
\includegraphics[angle=-90, width=0.325\textwidth]{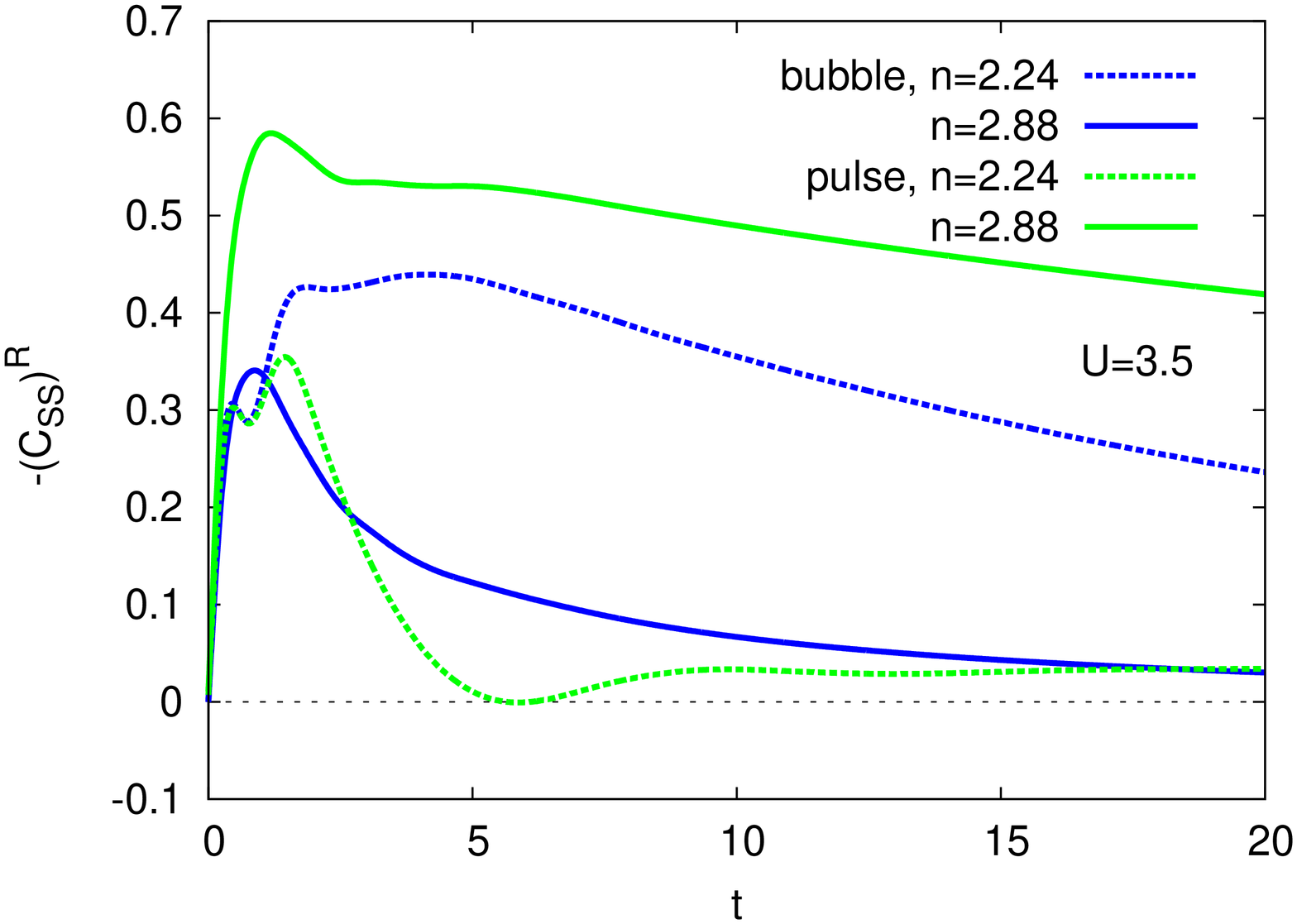}
\caption{
Left panel: NCA phase diagram for $\beta=50$ and $J=0.875$ in the intermediate-$U$ region. The pink line shows the boundary of the spin-triplet superconducting (SC) region and the blue line the location of the maximum superconducting order parameter. The physical region with repulsive interactions is above the dashed line ($U>3J$). 
Black bars indicate the Mott insulating solutions at filling $n=2$ and $n=3$.  
Middle panel: Filling dependence of the order parameter $O_P$ and of $\Delta\chi$ with (green) and without (blue) vertex corrections ($\beta=50$, $U=3.5$).
Right panel: Retarded spin-spin correlation functions evaluated with (green) and without (blue) vertex corrections in the underdoped and optimally doped regime ($\beta=50$, $U=3.5$).    
The spin correlation functions and susceptibilities are calculated in the normal phase. 
}
\label{fig_phasesc}
\end{center}
\end{figure*}

\section{Results}

\subsection{Equilibrium system}
\label{sec:equilibrium}

We start by presenting some results which demonstrate the spin-triplet pairing in equilibrium and its connection to local spin fluctuations.  The left panel of Fig.~\ref{fig_phasesc} shows the NCA phase diagram in the space of interaction $U$ and filling $n$ obtained for fixed $J=0.875$ at inverse temperature $\beta=50$. (The physical regime corresponds to $U>3J=2.625$, above the black dashed line.) We only plot the filling range $2\le n\le 4$, since the model is particle-hole symmetric. There is a Mott insulating solution at half-filling ($n=2$), and an $n=3$ Mott insulator for $U\ge U_c(n=3)\approx 5.875$, while a spin-triplet superconducting phase is found for $U \lesssim 4.1$ and fillings around $n=3$. The orbital-singlet spin-triplet superconducting order parameter $O_P$ 
at $U=3.5$ is plotted as a function of $n$ in the middle panel. It reaches its maximum near $n=2.93$. The blue line in the left panel tracks the maximum order parameter in the $U$-$n$ space. 
It connects to the end point of the $n=2$ Mott insulator, since for $U<U_c(n=2)$ (in the unphysical regime with attractive interorbital interactions) the highest $T_c$ is reached at half-filling. 
The appearance of this superconducting phase and its stability region in $U$ and $T$ space is qualitatively consistent with numerically exact data based on a QMC impurity solver.\cite{Werner2006Kondo,Hoshino2016}  
The main quantitative difference is that the superconducting instability is shifted to larger fillings in the NCA solution. 

\begin{figure*}[t]
\begin{center}
\includegraphics[angle=-90, width=0.325\textwidth]{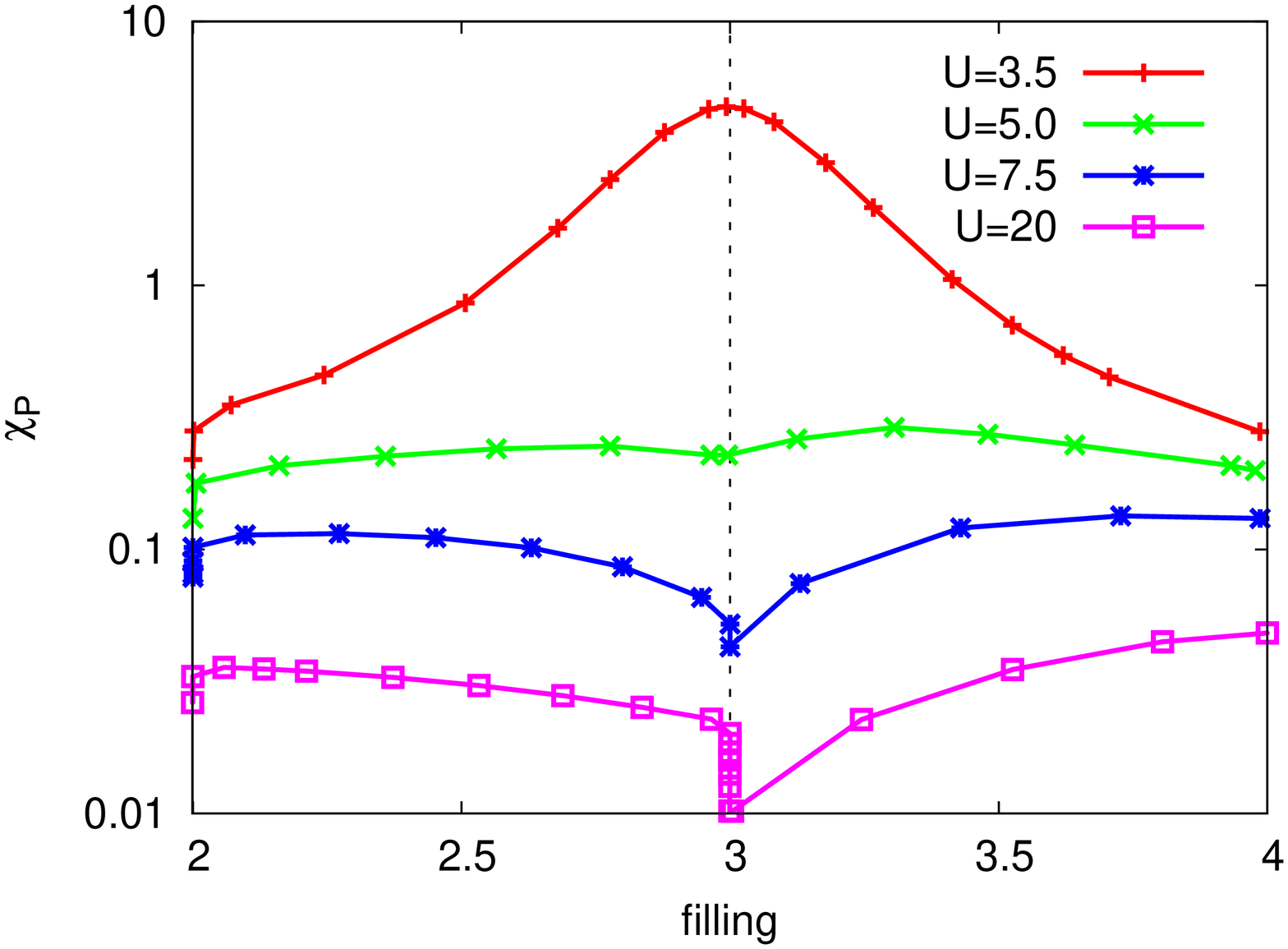}\hfill
\includegraphics[angle=-90, width=0.325\textwidth]{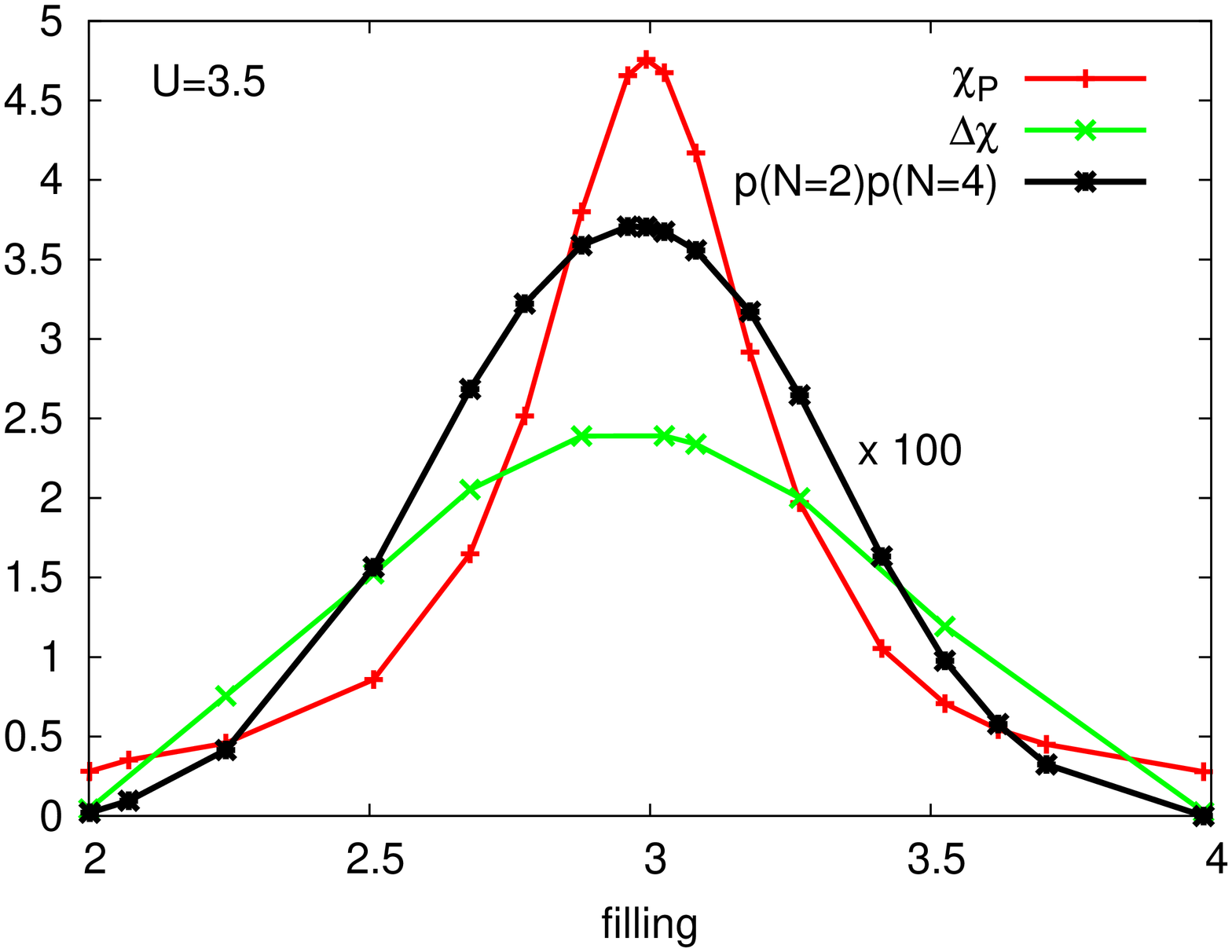}\hfill
\includegraphics[angle=-90, width=0.325\textwidth]{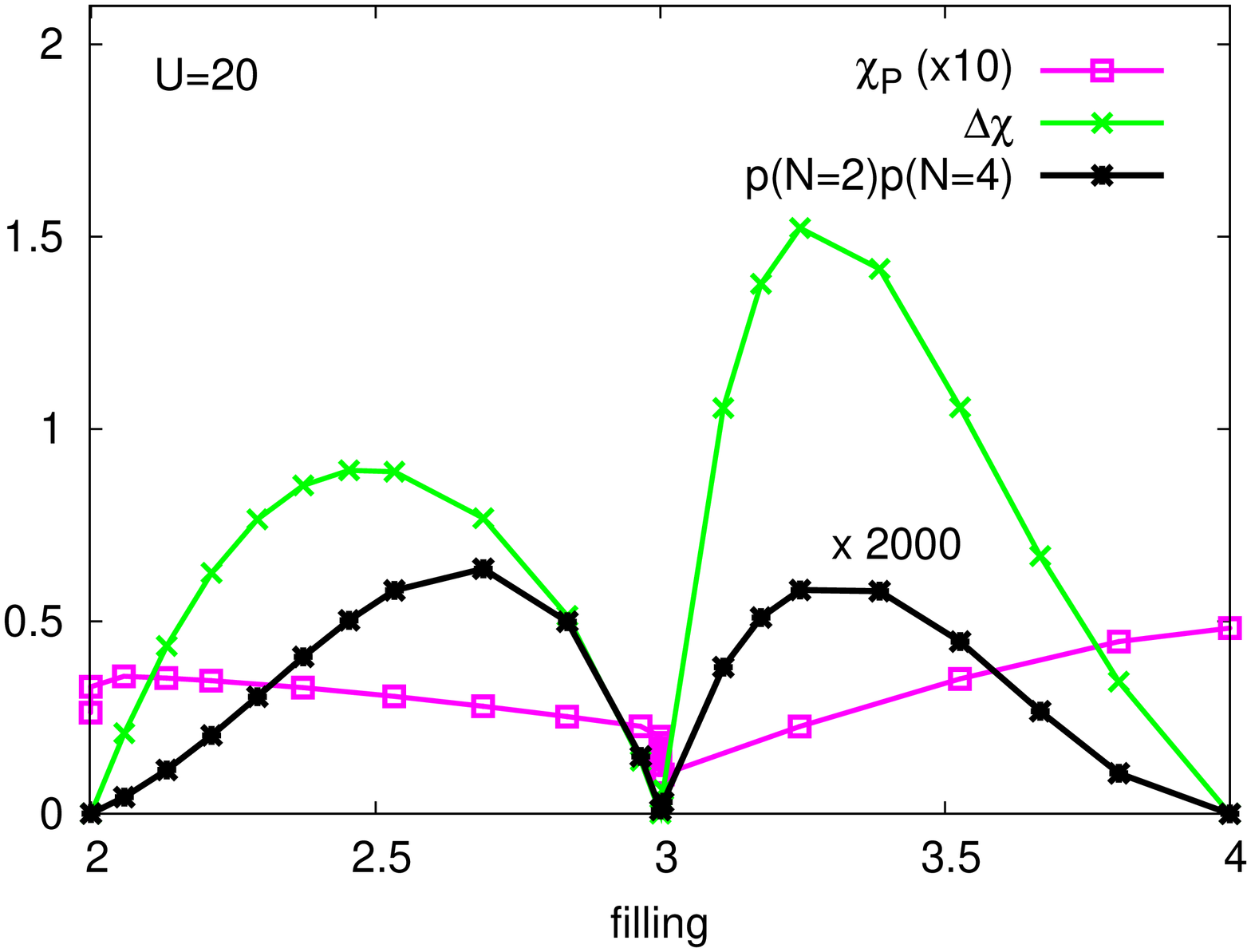}
\caption{
Equilibrium results for the chemically doped system at $\beta=25$ ($T>T_{c,\text{max}}$ in the considered interaction range). 
Left panel: Pairing susceptibility $\chi_P$ for different $U$ as a function of filling. For $U>U_c(n=3)\approx 5.875$ there exists an $n=3$ Mott insulator. 
Middle panel: Filling dependence of $\chi_P$ and the dynamical contribution to the local spin susceptibility $\Delta\chi$ for $U=3.5$.
Right panel: Filling dependence of $\chi_P$ and $\Delta\chi$ for $U=20$.
Also indicated is the product $p(N=2)p(N=4)$ of local state probabilities. 
}
\label{fig_sctime}
\end{center}
\end{figure*}      

In the QMC phase diagram,\cite{Hoshino2016} at intermediate $U$, the superconducting phase extends along the spin-freezing crossover line,\cite{Werner2008} where the dynamical contribution to the local spin-susceptibility $\Delta \chi$ reaches its maximum. As is shown in the middle panel of Fig.~\ref{fig_phasesc} (blue line), the NCA bubble approximation of $\Delta\chi$ peaks near $n=2.19$, outside of the filling range where the superconducting solution appears. This inconsistency is resolved by the magnetic field pulse measurement of the local spin susceptibility. This more accurate NCA estimate of $\Delta\chi$, which takes vertex corrections into account, is shown by the green line in the middle panel, and indeed exhibits a peak in the doping region where the superconducting order parameter reaches its maximum.  

\begin{figure*}[t]
\begin{center}
\includegraphics[angle=-90, width=0.32\textwidth]{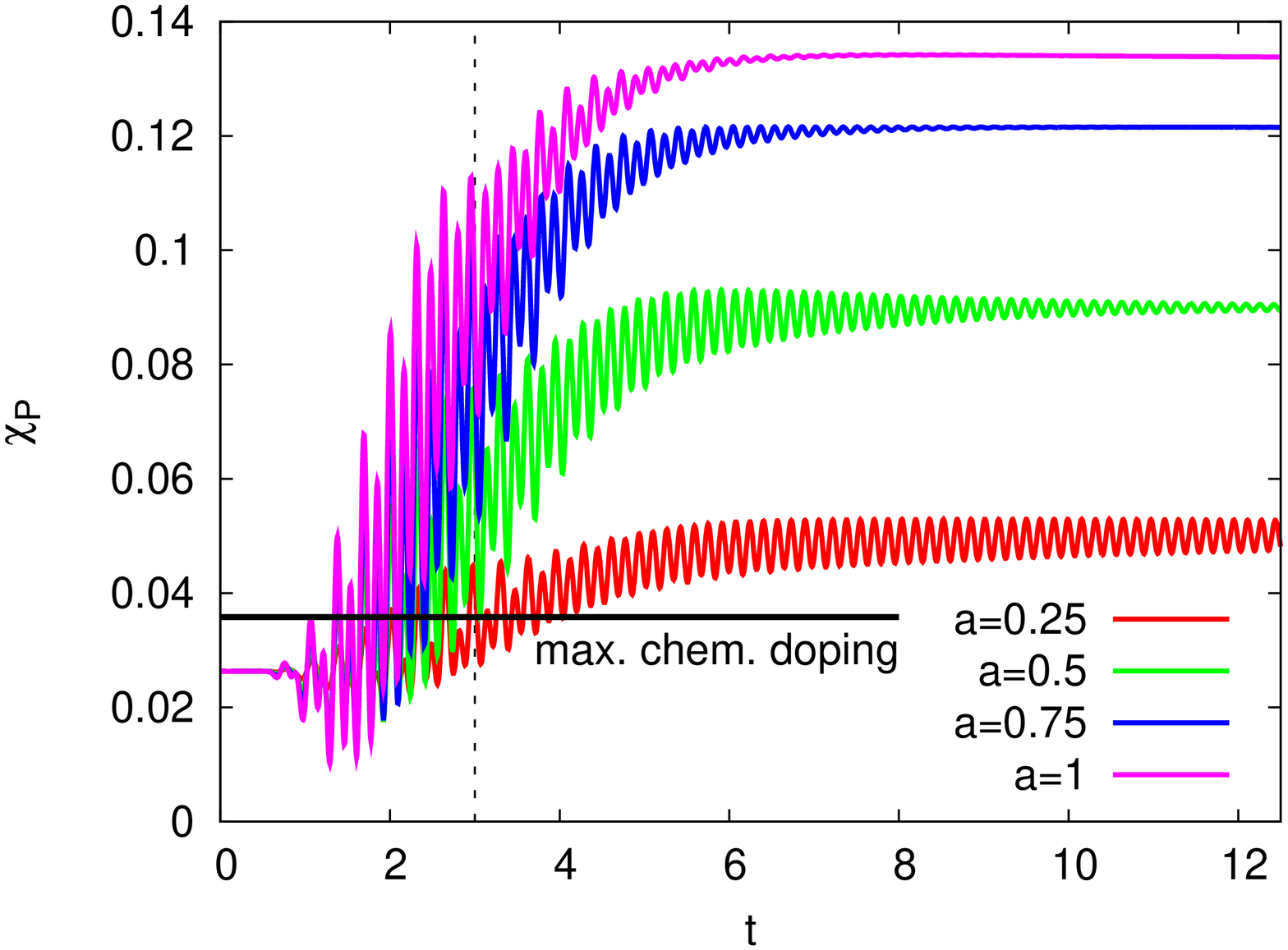}
\includegraphics[angle=-90, width=0.32\textwidth]{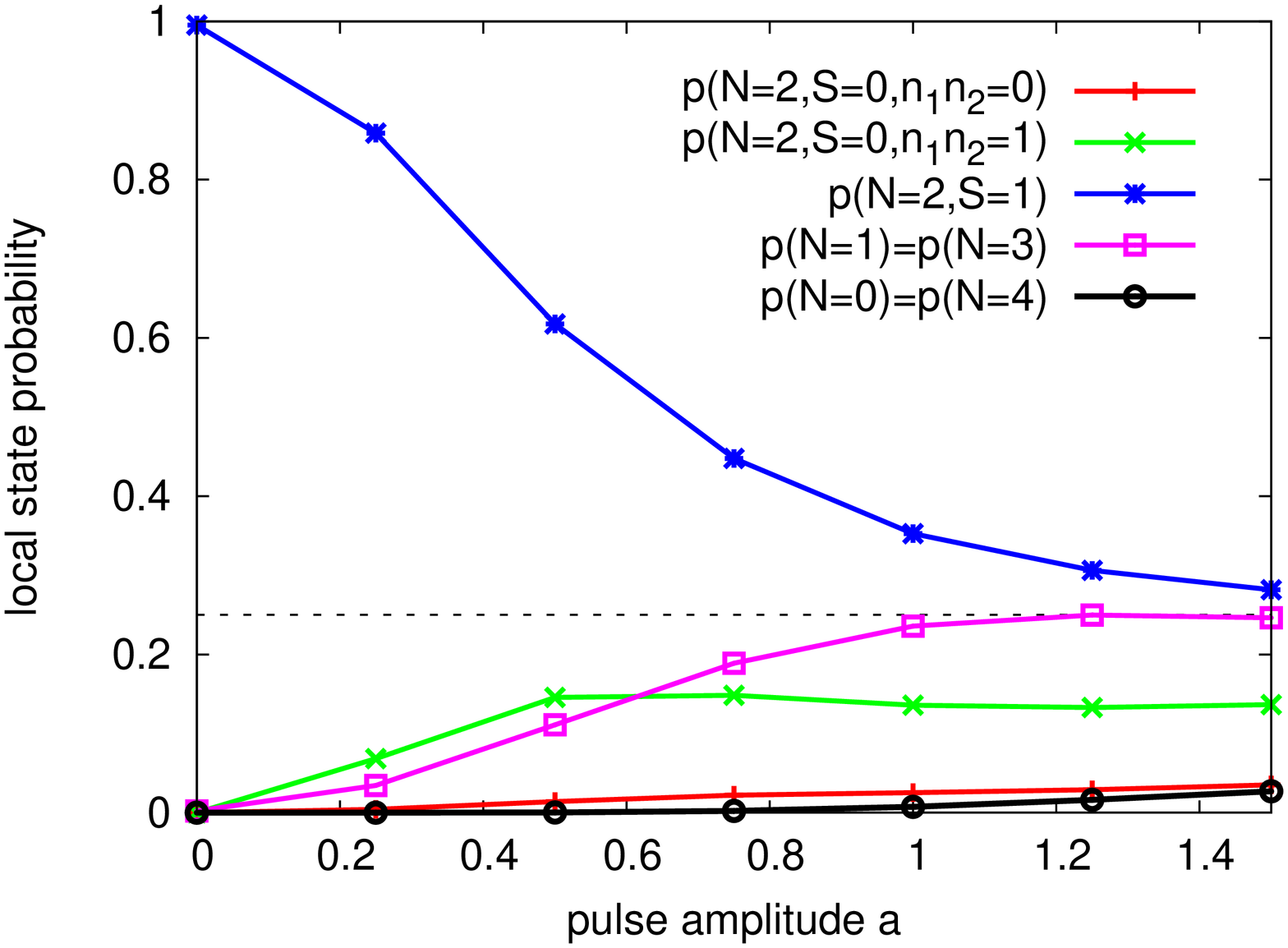}
\includegraphics[angle=-90, width=0.32\textwidth]{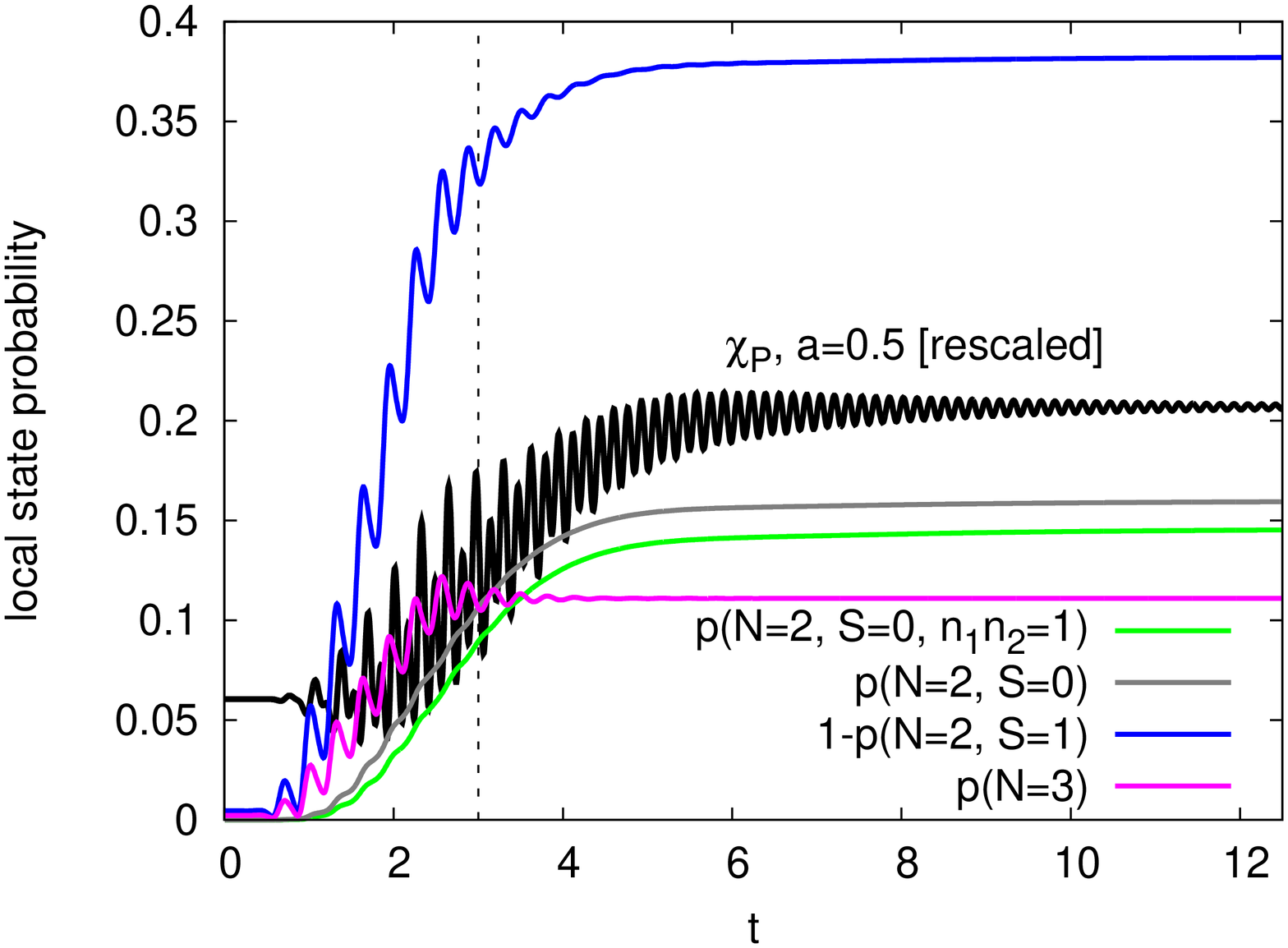}
\caption{
Photodoping of the half-filled Mott insulator with $U=20$, $J=0.875$, and initial $\beta=25$ by hopping modulation with amplitude $a$. Left panel: Pairing susceptibility $\chi_P$ measured by applying a constant seed field $P=0.001$. The horizontal black line shows the maximum pairing susceptibility that can be reached in the chemically doped system for fillings between $n=2$ and $n=3$ at $\beta=25$. The vertical dashed line indicates the end of the pulse. 
Middle panel: Distribution of local state probabilities at $t=12.5$ as a function of pulse amplitude $a$. ``$N=2,S=0,n_1n_2=0$" corresponds to the local states \{$|\!\uparrow\downarrow,0\rangle$,$|0,\uparrow\downarrow\rangle$\} and ``$N=2,S=0,n_1n_2=1$" to \{$|\!\uparrow,\downarrow\rangle$,$|\!\downarrow,\uparrow\rangle$\}. 
Right panel: Time evolution of the probabilities for different atomic states in the case of the pulse with amplitude $a=0.5$. 
}
\label{fig_photodoped}
\end{center}
\end{figure*}   

The large effect of the vertex corrections is illustrated for $U=3.5$ and two different fillings in the right hand panel of Fig.~\ref{fig_phasesc}, which plots the retarded spin-spin correlation function $C_{SS}^R(t)$ evaluated in the NCA bubble approximation (blue) and obtained by the pulse measurement according to Eq.~(\ref{eq_pulse}) (green). Near half-filling, in the spin-frozen regime, the bubble strongly overestimates the local spin correlations, while near optimal doping, in the spin-freezing crossover regime, it does not capture the slow decay of the spin correlations. The result obtained with the pulse measurement is qualitatively consistent with the susceptibility obtained on the imaginary axis by QMC [Eq.~(\ref{eq_delta_chi})].  
    
We next discuss the changes in the pairing susceptibility $\chi_P$ which occur as one moves into the large-$U$ regime. These changes are a consequence of the suppression of charge fluctuations near half-filling and the 
appearance of the $n=3$ Mott insulating phase at $U> U_c(n=3)\approx 5.875$.  
The left panel of Fig.~\ref{fig_sctime} plots $\chi_P$ at $\beta=25$ for different values of $U$. This temperature is above the maximum $T_c$ in the considered interaction range, and hence the pairing susceptibility does not diverge. As the system approaches $U_c(n=3)$, the peak in $\chi_P$ near $n=3$ gets suppressed, and for $U >  U_c(n=3)$ the susceptibility reaches a global minimum in the $n=3$ Mott insulating state. As a consequence, $\chi_P$ exhibits a broad maximum for some filling $n<3$, which shifts towards $n=2$ as $U$ is increased. At very large interaction, the NCA estimate of $\chi_P$ is strongly suppressed in the entire doping range $2<n<3$, and does not significantly exceed the values in the two bordering Mott phases. 

In the middle and right panels of Fig.~\ref{fig_sctime}, we compare the pairing susceptibility $\chi_P$ in the intermediate- and large-$U$ regimes to the dynamical contribution to the local spin susceptibility ($\Delta\chi$) evaluated by the pulse measurement (green line). For $U=3.5$ (middle panel) peaks in the pairing susceptibility and in $\Delta\chi$ are observed near $n=3$ in agreement with the previous discussion of the low-temperature phasediagram. However, for $U=20$ (right panel) the doping evolution of $\Delta\chi$ no longer correlates with the pairing susceptibility. While $\Delta \chi$ exhibits a pronounced maximum near $n=2.5$ and is suppressed by only about a factor of 2 compared to the $U=3.5$ case, the pairing susceptibility shows a very weak maximum near $n=2$ and its value is suppressed by a factor of 100 with respect to the $U=3.5$ result. 
This strong suppression of $\chi_P$ despite persistent spin fluctuations is consistent with QMC based DMFT simulations in the strong-coupling regime.
According to Ref.~\onlinecite{Hoshino2015}, the local spin fluctuations provide the ``pairing glue" for the orbital-singlet spin-triplet superconductivity. 
Hence, the peak in $\Delta\chi$ at $U=20$ indicates that the very low pairing susceptibility observed at $U=20$ is not due to a lack of pairing interactions, but due to  the strongly reduced charge fluctuations at large $U$ and the proximity of the chemically doped system with $2<n<3$ to the $n=2$ and $n=3$ Mott insulators.

\subsection{Photodoped system}
\label{sec:nonequilibrium}

We now move to the discussion of the nonthermal metal state obtained by photo-doping the half-filled Mott insulator. The photo-doping excitation is mimicked by a periodic modulation of the hopping parameter, with frequency $\Omega=U$, as explained in Sec.~\ref{sec:method}. This hopping modulation creates a certain density of triplons ($|N=3\rangle$) and singlons ($|N=1\rangle$) on top of a background of predominantly half-filled sites in a high-spin configuration ($|N=2, S=1\rangle$). In the large-gap regime, the life-time of these photo-excited charge carriers grows exponentially with increasing $U$,\cite{Strand2017,Eckstein2011} and for $U=20$ we can neglect the recombination of triplons and singlons on the timescales accessible in the simulations. 
We start by discussing the isolated system and then consider the effect of the coupling to a boson bath. 

\subsubsection{Isolated system}

To illustrate the evolution of the pairing susceptibility, we apply a small constant pair seed-field $P(t)=p=0.001$ and calculate the order parameter $O_P(t)$. This yields the estimate $\chi_P(t)=O_P(t)/p$. The left panel of Fig.~\ref{fig_photodoped} shows the result for different modulation amplitudes $a$ and a pulse which lasts from $t\approx 0.5$ to $t\approx 3$. Also plotted for comparison is the largest value of the pairing susceptibility $\chi_P$ that can be reached in equilibrium at $\beta=25$ in the filling range $2\le n \le 3$. The photo-doping leads to a prompt increase of $\chi_P$ and the effect grows with the amplitude of the pulse.  
Even though the pulse injects energy, and the pairing susceptibility of the equilibrium system decreases with increasing temperature, we find that the photo-doped system reaches pairing susceptibilities that are substantially larger than in equilibrium, and that the effect is long-lived in the absence of singlon-triplon recombination. 

In the middle panel we plot the local state probabilities as a function of pulse amplitude $a$ in the steady state reached after the pulse ($t=12.5$). The main effect of the pulse is to transform high-spin doublon states into singlons and triplons. At small pulse amplitudes, there is also a significant increase in the density of low-spin doublon states of the type $|\!\uparrow,\downarrow\rangle$  or $|\!\downarrow,\uparrow\rangle$ (``$n_1n_2=1$") with increasing $a$. At large pulse amplitudes, the singlon and triplon density saturates at 0.25. A further increase in the pulse amplitude or the length of the pulse then mainly leads to the generation of $|N=0\rangle$ and $|N=4\rangle$ states at the expense of the remaining high-spin doublons. 

One remarkable observation is that the pairing susceptibility continues to increase after the end of the pulse (between $t\approx 3$ and $t\approx 6$). To understand this behavior, let us look at the time evolution of the probabilities of the different atomic states. For the pulse with amplitude $a=0.5$ the result is shown in the right panel of Fig.~\ref{fig_photodoped}. The pink line shows the probability of a given site to be in a triplon state ($p(N=3)$, which by symmetry is equal to the probability of the singlon state $p(N=1)$), while the blue curve plots the probability of not being in the high-spin doublon state ($1-p(N=2, S=1)$). We can see that the number of triplons is indeed conserved after the pulse, up to some rapidly damped oscillations associated with $(N=1) (N=3)\leftrightarrow (N=2, S=1)( N=2, S=1)$ hopping processes, which also lead to strong oscillations in the kinetic energy. During the pulse, the probability of the high-spin doublon state is reduced, while the occupation of singlon, triplon, and low-spin doublon states grows. The increase of the pairing susceptibility after the pulse is associated with a further transformation of high-spin doublon states into low-spin doublon states at fixed density of singlons and triplons. This is a clear signature of the cooling of the photo-doped singlons and triplons by local spin excitations, similarly to what has been demonstrated in Ref.~\onlinecite{Strand2017}. This cooling is also evident in the time-evolution of the kinetic energy, 
which decreases after the pulse on the 
same timescale as the observed transformation of high-spin doublon states into low-spin doublon states. 

\begin{figure}[t]
\begin{center}
\hspace{-0.5mm}\includegraphics[angle=-90, width=0.81\columnwidth]{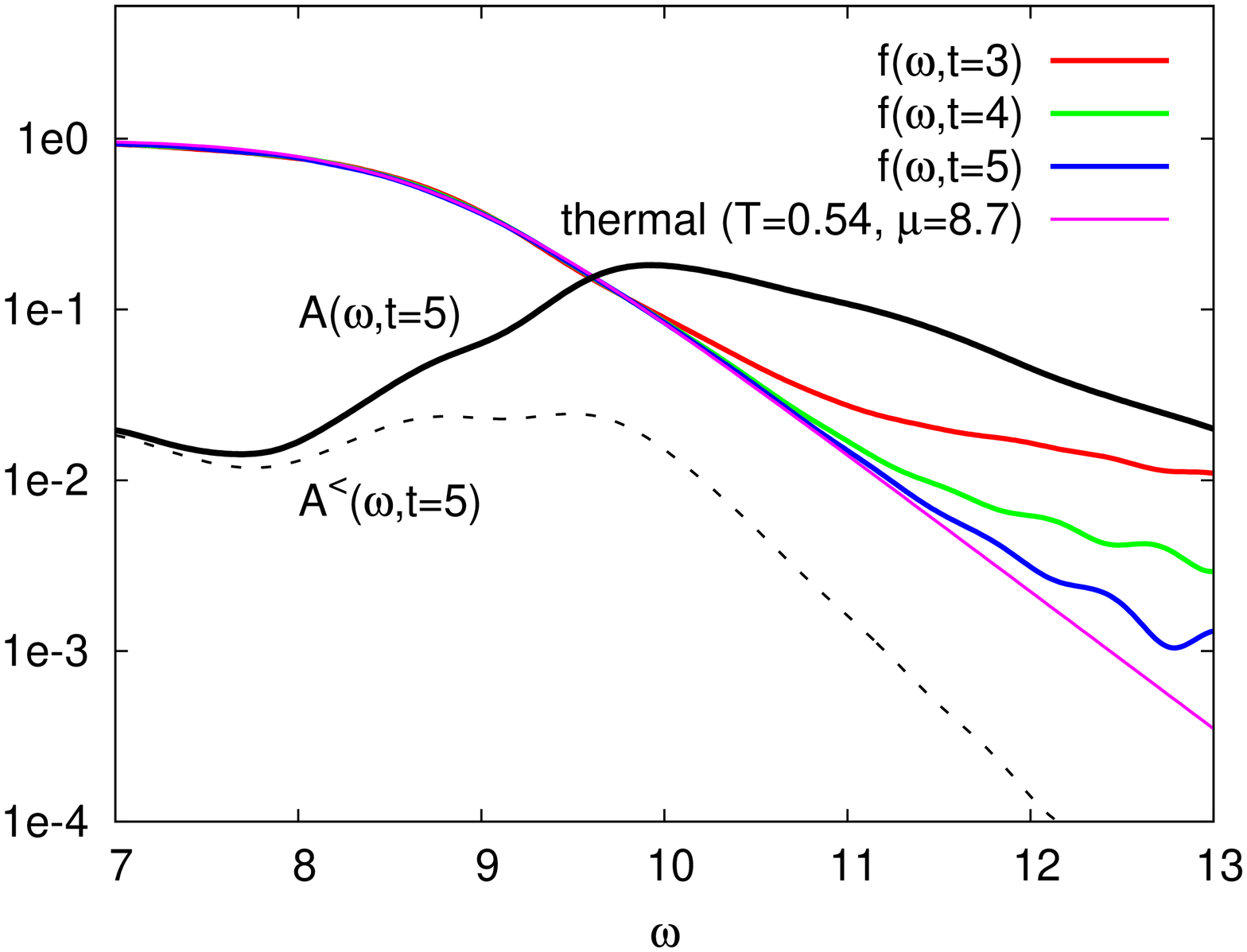}\\
\includegraphics[angle=-90, width=0.8\columnwidth]{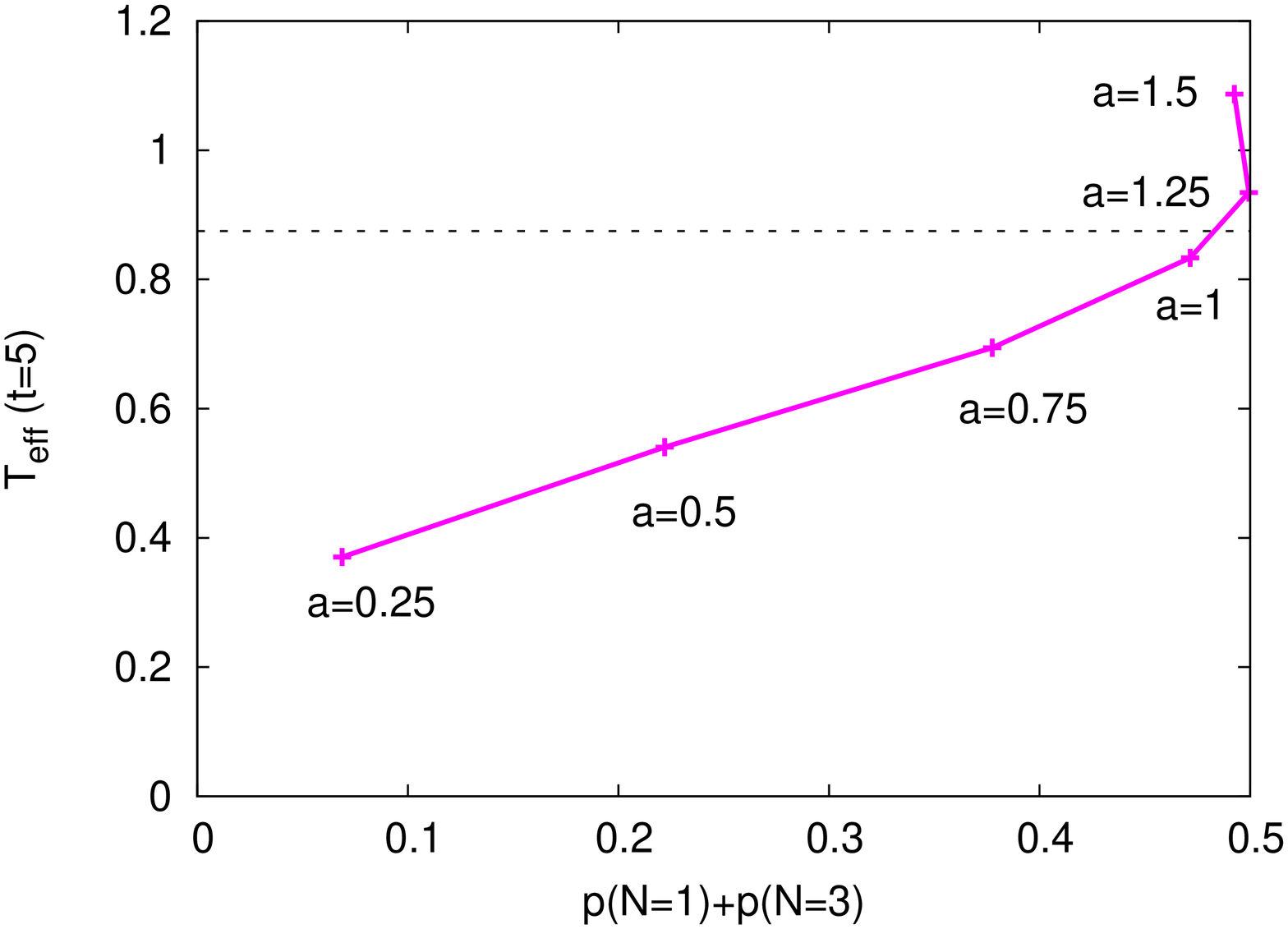}
\caption{
Effective triplon temperature in the photo-doped system. Top panel: Time-dependent energy distribution function $f(\omega,t)$ for $U=20$, $J=0.875$, initial $\beta=25$ and pulse amplitude $a=0.5$, and comparison to a Fermi-Dirac distribution function with $T=0.54$ and $\mu=8.7$. 
Bottom panel: Effective triplon temperature at time $t=5$ plotted as a function of the photo-doping concentration $p(N=1)+p(N=3)$. The dashed horizontal line indicates the value of $J$.
}
\label{fig_teff_triplon}
\end{center}
\end{figure} 

By measuring the time-dependent spectral function of the 
photo-doped state as the Fourier transform of the retarded component of the local 
Green's function,\cite{footnote_component} 
$A(\omega,t)=-\frac{1}{\pi}\text{Im}\int_t^{t_\text{max}} dt' e^{i\omega(t'-t)}G_{11}^\text{R}(t',t)$ and the occupation function  $A^<(\omega,t)=\frac{1}{2\pi}\text{Im}\int_t^{t_\text{max}} dt' e^{i\omega(t'-t)}G_{11}^<(t',t)$ 
as the Fourier transform of the lesser component, we can define the nonequilibrium distribution function $f(\omega,t)=A^<(\omega,t)/A(\omega,t)$. 
The top panel of Fig.~\ref{fig_teff_triplon} plots $f(\omega,t)$ in the energy range of the upper Hubbard band, for times immediately after the hopping modulation pulse with amplitude $a=0.5$. In contrast to the paramagnetic single-orbital model, where a nonthermal distribution of photo-carriers persists for a long time after a photo-doping pulse\cite{Eckstein2011} we observe a relaxation of the high-energy triplons to an approximate Fermi-Dirac distribution with a shifted chemical potential on the timescale of the inverse hopping. The effective temperature of the photo-doped triplons depends on the pulse amplitude (bottom panel of Fig.~\ref{fig_teff_triplon}) or ``photo-doping concentration" $p(N=1)+p(N=3)$, but it is of the order of the Hund coupling $J$. The evolution of the nonequilibrium distribution function thus provides further evidence for the ultra-fast cooling of the photo-doped triplons and singlons by local spin excitations, to an effective temperature of the order of $J$. (Additional cooling is possible if energy quanta smaller than $J$ can be dissipated to some heat bath, see Sec.~\ref{sec:bath}.) As the pulse amplitude reaches $a\approx 1$, the high-spin doublon states are depleted (Fig.~\ref{fig_photodoped}) and local spin excitations with energy cost $J$ become rare. In this regime the effective temperature grows beyond $T_\text{eff}=J$ with increasing $a$ (or increasing length of the pulse), while the photo-doping concentration saturates near $0.5$. 

\begin{figure}[t]
\begin{center}
\includegraphics[angle=-90, width=0.8\columnwidth]{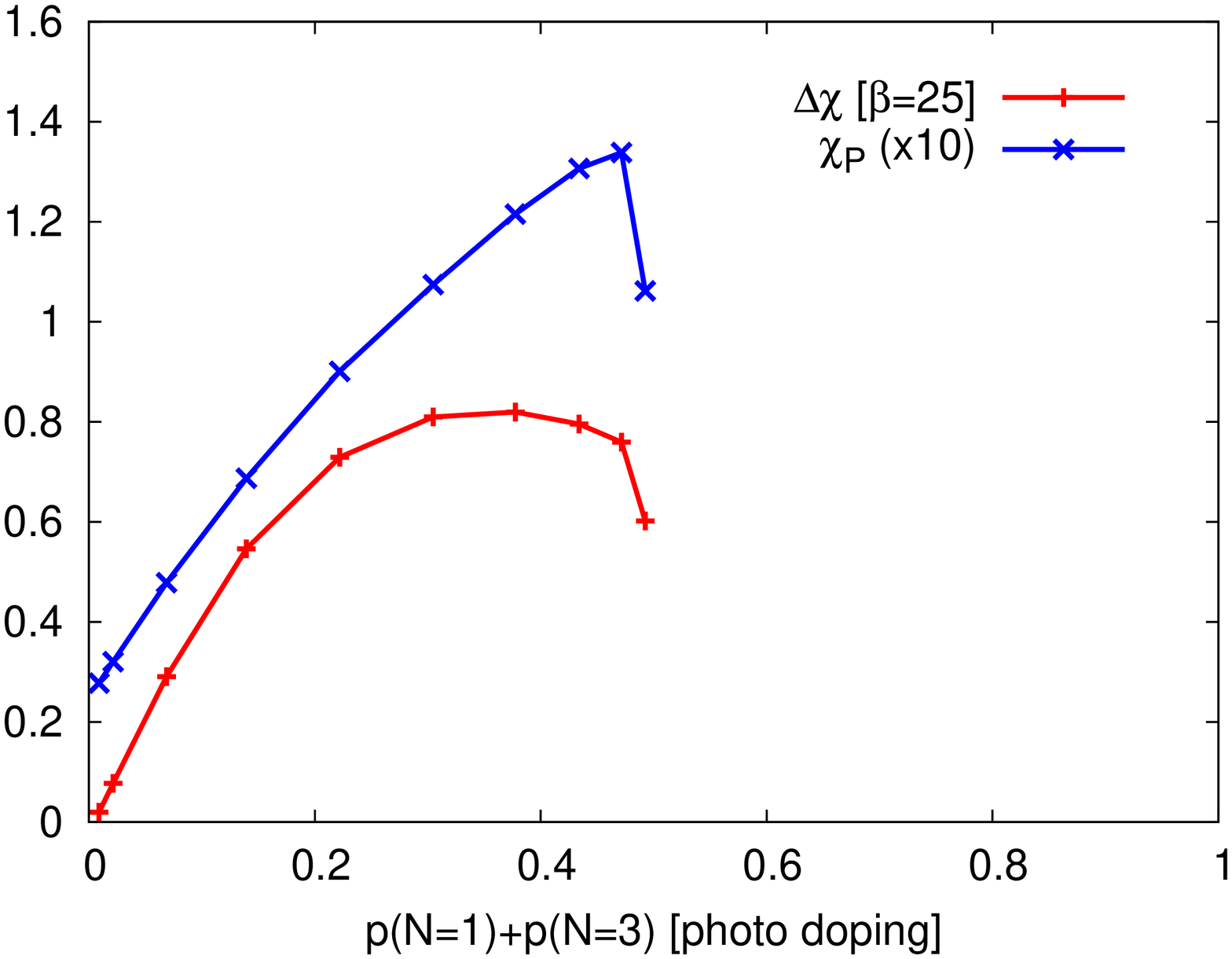}\\
\includegraphics[angle=-90, width=0.8\columnwidth]{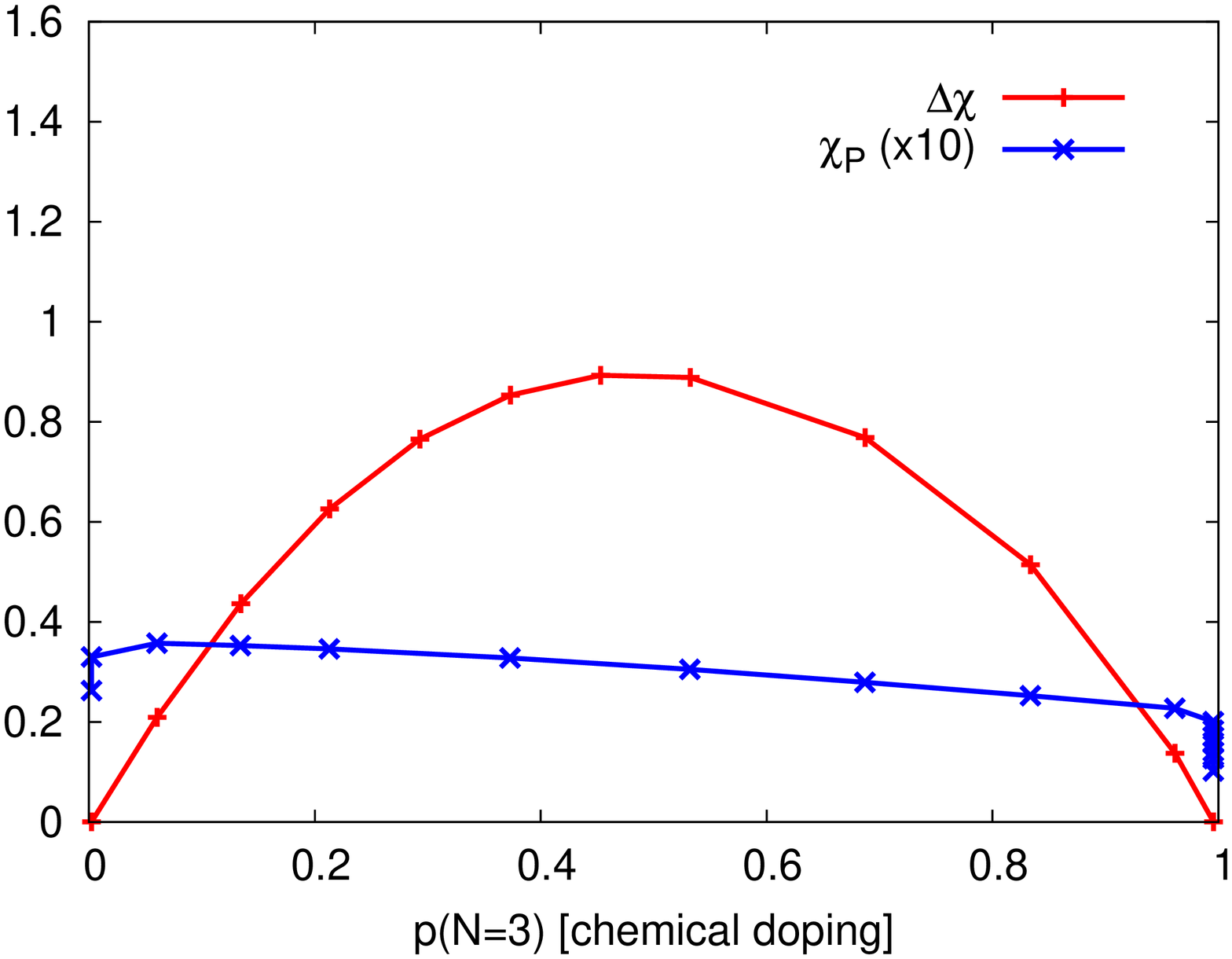}
\caption{
Susceptibilities in the photo-doped and chemically doped system with $U=20, J=0.875.$ 
Top panel: Pairing susceptibility $\chi_P$ in the photo-doped system plotted against the ``photo-doping concentration" $p(N=1)+p(N=3)$, and dynamical contribution to the local spin susceptibility $\Delta\chi$ evaluated according to Eq.~(\ref{eq_pulse}) with $\beta=25$. 
Bottom panel: $\chi_P$ and $\Delta\chi$ of the chemically doped equilibrium system at $\beta=25$ plotted against the ``doping concentration" $p(N=3)$. 
}
\label{fig_susc_triplon}
\end{center}
\end{figure}   

To compare the susceptibility of the photo-doped system to that of a chemically doped equilibrium system at inverse temperature $\beta=25$ we plot in Fig.~\ref{fig_susc_triplon} the pairing susceptibility and the local spin susceptibility as a function of carrier concentration ($p(N=1)+p(N=3)$ in the photo-doped case and $p(N=3)$ in the chemically doped case). 
While the chemically doped system at $U=20$ shows a very weak maximum in $\chi_P$ near $n=2$, there is a clear increase with increasing photo-doping concentration in the nonequilibrium state. The effect saturates near $p(N=1)+p(N=3)=0.5$, since this corresponds to the largest singlon/triplon density that can be reached using a pulse excitation with $\Omega=U$. As mentioned above, the cooling by local spin fluctuations becomes less effective as this limit is approached. 

We also plot the dynamical contribution to the local susceptibility, evaluated with Eq.~(\ref{eq_pulse}) and $\beta=25$
in the quasi steady-state reached after the pulse $(t_p=6)$. Since the global temperature of the nonequilibrium state is not defined, it is a priori not clear how to apply this formula, and by inserting the ``effective temperature" $1/\beta\approx J$ of the triplons, one would obtain much smaller values of $\Delta\chi$. By using the low temperature of the initial state, Eq.~(\ref{eq_pulse}) essentially measures the long-time decay of the retarded spin-spin correlation function, and Fig.~\ref{fig_susc_triplon} thus allows us to compare these decays in the chemically doped and photo-doped states. The result is rather similar, apart from the reduction near $p(N=1)+p(N=3)=0.5$ in the photo-doped state, which can be assigned to heating effects.

Figure~\ref{fig_susc_triplon} demonstrates a correlation between $\chi_P$ in the photo-doped state at $U=20$ and $\Delta\chi$, which is qualitatively similar to what is found in equilibrium in the moderately correlated regime ($U< U_c(n=3)$), see middle panel of Fig.~\ref{fig_sctime}. This indicates that the build-up of coherence in this photo-doped 
metastable state is connected to local spin fluctuations in much the same way as was found for the moderately correlated, chemically doped equilibrium system. 

Let us comment at this point on the functional dependence of $\chi_P$ on the doping concentration. In equilibrium, at $U=3.5$, the pairing susceptibility peaks near $n=3$ and it correlates with the product $p(N=2)p(N=4)$ of local state probabilities (Fig.~\ref{fig_sctime}). This is natural, since in an equilibrium system with three electrons per site, the triplet pairing is associated with fluctuations between $N=2$ and $N=4$ states. In the photo-doped half-filled Mott insulator, the singlons and triplons move in a background of predominantly high-spin $N=2$ states. Hence, the analogy to the moderately correlated equilibrium superconductor suggests a correlation between $\chi_P$ and $p(N=1)p(N=3)=p(N=3)^2$ in the photo-doped system. The top panel of  Fig.~\ref{fig_susc_triplon} shows a different scaling (very roughly $\chi_P\sim p(N=3)^{1/2}$). We argue that this is a consequence of the doping-dependent effective temperature of the photo-doped system, and will come back to this point at the end of the next section, which discusses the effect of cooling by a bosonic heat bath.

\subsubsection{System coupled to a heat bath}
\label{sec:bath}

We finally consider the evolution of the photo-doped state in the presence of a bosonic heat bath. As discussed in the previous subsection, local spin excitations provide a very efficient cooling mechanism for the photo-doped triplons and singlons, so that only a few inverse hopping times after the pulse, the effective temperature of these carriers is of the order of $J$. Below this effective temperature, 
the kinetic energy of the singlons and triplons is too low to excite high-spin states into low-spin states, so that the intra-band relaxation is limited by an effect analogous to the ``phonon bottleneck" discussed in the context of photo-excited electron-phonon systems.\cite{Sentef2013,Murakami2015}
In the presence of a bosonic heat-bath described by the self-energy (\ref{eq_bath}) with boson energy $\omega_0<J$, further cooling of the singlons and triplons is possible, on a timescale determined by 
$\omega_0$ and the coupling strength $g$, and it is an interesting question how this cooling affects the pair susceptibility of the photo-doped state.   

\begin{figure*}[t]
\begin{center}
\includegraphics[angle=-90, width=0.325\textwidth]{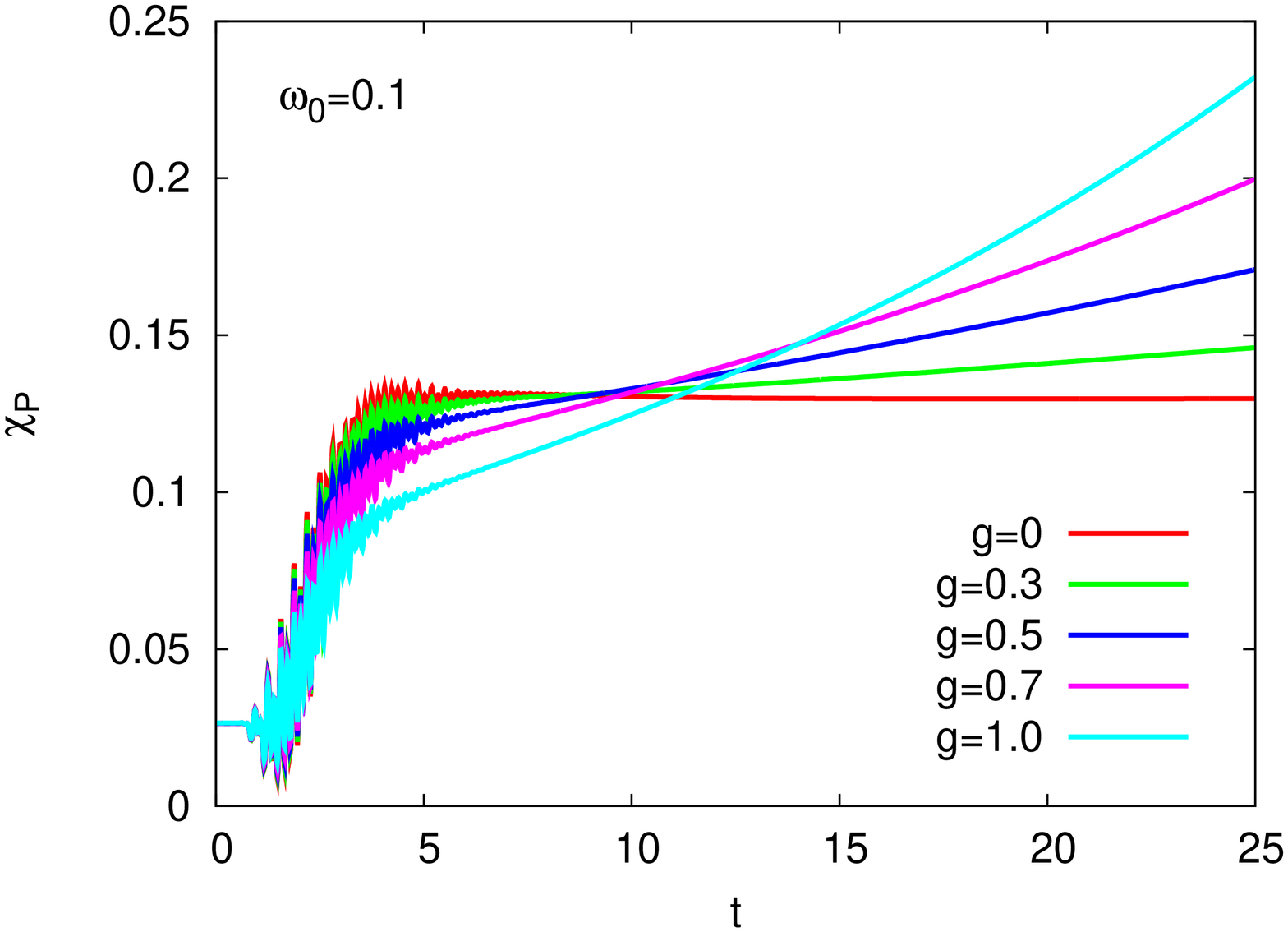}\hfill
\includegraphics[angle=-90, width=0.325\textwidth]{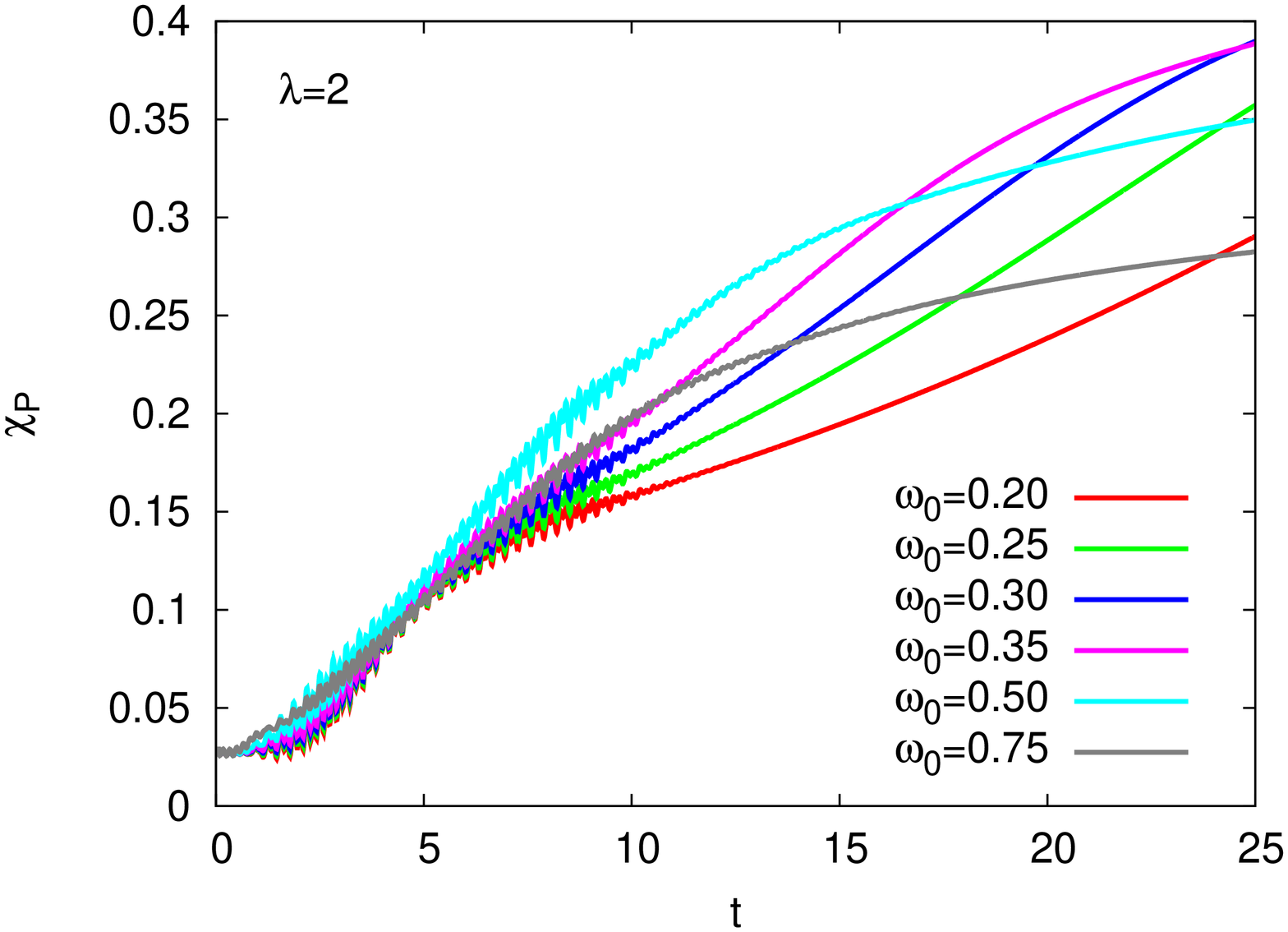}\hfill
\includegraphics[angle=-90, width=0.325\textwidth]{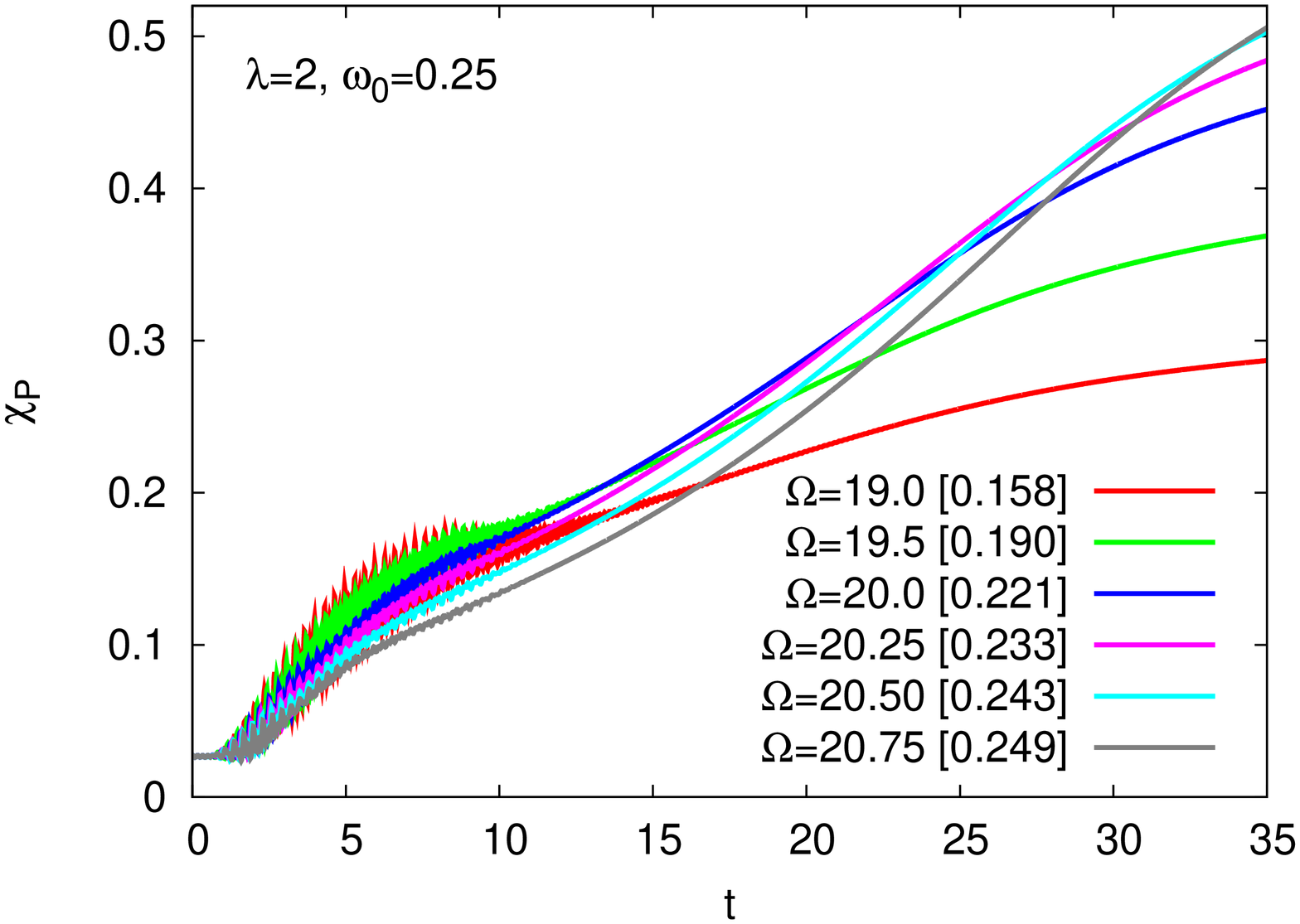}
\caption{
Time evolution of the pairing susceptibility in a photo-doped Mott insulator with coupling to a bosonic heat bath ($U=20$, $J=0.875$, initial $\beta=25$). 
Left panel: Results for fixed $\omega_0=0.1$ and different couplings $g$ (short pulse with amplitude $a=1$). 
Middle panel: Results for fixed effective boson coupling $\lambda=2$ and different phonon frequencies for a longer pulse with amplitude $a=0.5$, which essentially saturates the number of triplons ($p(N=3)\approx 0.22$ after the pulse, except for $\omega_0=0.75$). 
Right panel: Results for fixed $\lambda=2$, $\omega_0=0.25$, pulse amplitude $a=0.5$ and different pulse frequencies $\Omega$. The number between brackets is the density of triplon states $p(N=3)$ after the pulse. 
}
\label{fig_cooling}
\end{center}
\end{figure*}   

In the left panel of Fig.~\ref{fig_cooling} we show the time evolution of the pairing susceptibility for pulse amplitude $a=1$,  boson frequency $\omega_0=0.1$, and different coupling strengths $g$. First of all we note that in equilibrium, the coupling to a boson bath decreases the pairing susceptibility relative to the model without heat bath (although the effect is very small in the half-filled Mott insulator). In the photo-doped state, however, the additional cooling by the boson bath results in a further enhancement of the pairing susceptibility. While $\chi_P$ immediately after the pulse is reduced compared to the system without heat-bath (a result of the detrimental effect of the bath self-energy, and changes in the photo-doping concentration), a strong boson coupling results in a substantial increase of the pairing susceptibility at later times. The growth rate is faster for larger $g$, because the energy dissipation rate of photo-carriers is determined by $\lambda=2g^2/\omega_0$ if the initial kinetic energy is much larger than $\omega_0$.\cite{Werner2015}

In the middle panel of Fig.~\ref{fig_cooling} we show results for a fixed $\lambda= 2$ and different boson frequencies. (Here, the pulses are longer, but their amplitude is reduced to $a=0.5$.) The density of triplons in the photo-doped state decreases slightly from 0.226 to 0.210 as $\omega_0$ is increased from 0.2 to 0.5 and then there is a substantial drop to 0.171 for $\omega_0=0.75$. Despite this trend, the pairing susceptibility initially increases substantially with increasing boson frequency, which implies a more efficient cooling. 
Hence, in a situation where the effective temperature is already reduced by local spin excitations to a value of the order of $J=0.875$ and $\omega_0$ is not much smaller than this energy scale, the dynamics depends explicitly on $\omega_0$, even for fixed effective coupling $\lambda$. 
As the boson frequency becomes comparable to $J$ the additional cooling by the boson bath becomes more and more limited, which manifests itself in a saturation of the susceptibility. On the numerically accessible timescales, the largest pairing susceptibility is reached for $\omega_0 \approx 0.3$, but we expect that smaller boson frequencies will lead to an even bigger enhancement at later times. 

The right hand panel of Fig.~\ref{fig_cooling} plots the evolution of the pairing susceptibility for fixed $\lambda=2$, $\omega_0=0.25$ and a long pulse with amplitude $a=0.5$, with pulse frequency $\Omega$ varying in the range $19\le \Omega\le 21$. Also indicated is the density of triplons in the photo-doped state. This data set confirms that for fixed $\lambda$, the largest pairing susceptibility is reached if the triplon density is saturated at $p(N=3)=0.25$. 

\begin{figure}[t]
\begin{center}
\includegraphics[angle=-90, width=0.8\columnwidth]{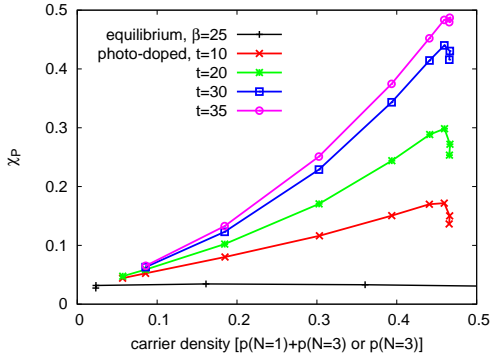}
\caption{Pairing susceptibility $\chi_P$ as a function of $p(N=3)$ in the equilibrium system with $\beta=25$, and as a function of $p(N=1)+p(N=3)$ in the photo-doped system at indicated values of the time $t$. The parameters of the boson bath are $\lambda=2$, $\omega_0=0.25$ and $\beta=25$.
}
\label{fig_cooling_doping}
\end{center}
\end{figure}

We finally plot in Fig.~\ref{fig_cooling_doping} a comparison between the doping-dependent pairing susceptibility in equilibrium and in the photo-doped metal state at different times. The bath parameters are $\lambda=2$ and $\omega_0=0.25$ and the inverse temperature of the initial state and the boson bath is $\beta=25$. The pairing susceptibility in the equilibrium state is about 0.027, similar to the case without heat bath (see Fig.~\ref{fig_susc_triplon}). In the photo-doped state, $\chi_P$ increases with the density of carriers $p(N=1)+p(N=3)=2p(N=3)$ and with increasing time (cooling of the carriers). With the chosen bath parameters, we observe an approximately 20-fold increase of the pairing susceptibility in the strongly photo-doped system on the accessible time-scales.

In contrast to the isolated system, which exhibited an unusual scaling of $\chi_P$ with $p(N=3)$, as illustrated in the top panel of Fig.~\ref{fig_susc_triplon}, the photo-doped system with boson bath exhibits an approximately quadratic dependence of  $\chi_P$ on $p(N=3)$ at later times. This shows that the pairing susceptibility in this cooled photo-doped Mott insulator correlates with the product $p(N=1)p(N=3)$. Since spin-triplet pairing in a half-filled system is associated with fluctuations between $N=1$ and $N=3$ states, this indicates that populating these charge states to overcome the suppression of charge fluctuations by $U$ helps superconductivity. We thus find that the photo-doped large-$U$ insulator exhibits a correlation between the pairing susceptibility $\chi_P$, the dynamical contribution to the local spin susceptibility $\Delta\chi$ and the product of state probabilities $p(N=1)p(N=3)$, which is qualitatively different from the chemically doped $U=20$ insulator, but analogous to the moderately correlated equilibrium system at $U<U_{c}(n=3)$. In the latter case the pairing susceptibility peaks near three electron filling, and correlates with the doping evolution of $\Delta\chi$ and $p(N=2)p(N=4)$.

\section{Summary and Conclusions}
\label{sec:summary}

We studied the spin-triplet pairing susceptibility in a photo-doped two-orbital Hubbard model using nonequilibrium DMFT in combination with an NCA impurity solver. Multi-orbital Hubbard models with $J>0$ exhibit an orbital-singlet spin-triplet superconducting phase if one dopes the half-filled insulator at low temperature. The pairing is induced by slowly fluctuating local moments in the spin-freezing crossover regime, as evidenced by the close correlation between the maximum $T_c$ (or maximum pairing susceptibility $\chi_P$) and the maximum in the dynamical contribution to the local spin susceptibility $\Delta\chi$.\cite{Hoshino2015,Hoshino2016} This correlation holds up to $U\lesssim U_c(n=3)$, whereas the appearance of the $n=3$ Mott insulator and the suppression of charge fluctuations at larger interactions leads to a strongly reduced $T_c$ in the filling range $2<n<3$.

While the NCA approximation has clear limitations in the application to multi-orbital systems, we found that this simple impurity solver, combined with a magnetic field-pulse measurement of the local spin susceptibility, captures the main qualitative features of the equilibrium DMFT phase diagram, such as the orbital-singlet spin-triplet superconducting instability in the doped Mott insulator, the correlation between $\chi_P$ and $\Delta\chi$ for $U\lesssim U_c(n=3)$ and the strong suppression of $\chi_P$ for $U > U_c(n=3)$ in the filling range $2<n<3$. Since perturbative strong-coupling expansions are at present the only available methods for the real-time simulation of multi-orbital impurity models in the strong correlation regime, and the numerically more expensive one-crossing approximation\cite{Pruschke1989} does not necessarily result in a qualitative improvement,\cite{Strand2017} we used the NCA solver in our study of the nonequilibrium pairing susceptibility of the photo-doped two-orbital model. 

In the moderately correlated regime $U<U_c(n=3)$, photo-excitation with pulse frequency $\Omega=U$ leads to a rapid heating of the system and a corresponding suppression of the pairing susceptibility. The reason is a pronounced change in the spectral function of a photo-doped multi-orbital model: a non-negligible density of singlon and triplon states leads to sidebands split off by an energy $\approx 3J$, which substantially reduce or completely fill the gap. 
This enables a fast recombination of singlons and triplons in the intermediate-$U$ regime and results in a strong heating. 
We thus focused our study on the large-$U$ regime, where the gap size is so large that the density of photo-doped singlons and triplons is approximately conserved on the numerically accessible timescales. In this case, a large part of the injected energy is stored as potential energy, which enables the emergence of interesting ``low-temperature" quantum phenomena such as superconductivity.

The photo-doped large-$U$ Mott insulator represents a genuine nonequilibrium state of matter, with properties that are distinct from those of a chemically doped Mott insulator. The most obvious difference is the nature of the charge carriers. In a chemically doped Mott insulator with an average density $2<n<3$, the charge carriers are predominantly triplons. If spin triplet pairing occurs near density $n=3$, this pairing is associated with fluctuations between local $N=2$ and $N=4$ states. At $U>U_c(n=3)$ the chemically doped state turns into a Mott insulator and the proximity to this insulating phase leads to a strong suppression of the conductivity and pairing susceptibility near filling $n=3$. In the photo-doped $n=2$ insulator, the charge carriers are singlons and triplons moving in a background of predominantly high-spin doublon states. If spin-triplet pairing occurs in such a system, it is associated with fluctuations between local $N=1$ and $N=3$ states. 
Hence, producing singlons and triplons by photo-doping may be a way to overcome the suppression of charge fluctuations by $U$ and to enable a build-up of coherence. In practice, we can reach an effective doping concentration $p(N=1)+p(N=3)=0.5$ with resonant pulses ($\Omega=U$). 

An interesting point is the connection between spin-triplet pairing and local spin fluctuations. 
While the close correlation between $\chi_P$ and $\Delta\chi$ is lost in the large-$U$ chemically doped system, it is recovered in the photo-doped half-filled Mott insulator. 
The pairing susceptibility in the cooled metastable state is connected both to the local spin susceptibility and the product of local state probabilities $p(N=1)p(N=3)$ and thus behaves in an analogous way to the chemically doped intermediate-$U$ model, which exhibits the same type of correlations between $\chi_P$, $\Delta\chi$ and $p(N=2)p(N=4)$ near filling $n=3$. The unconventional pairing mechanism revealed in Ref.~\onlinecite{Hoshino2015} thus appears to be active also in the photo-doped nonequilibrium state. Whether or not there are additional pairing channels in this large-$U$ system is an interesting question for further investigations.

While we have focused in this work on spin-triplet pairing and its relation to local spin fluctuations, it would be  worthwhile to extend the  study to intra-orbital pairing and to the photo-doped single-orbital case. The singlon and triplon states in the photo-doped two-orbital model also enhance the mobility of intra-orbital spin-singlet pairs. We may thus expect an enhanced spin-singlet pairing susceptibility in the photo-doped half-filled system, which has no analogy to the chemically doped case near $n=3$. (In the latter system the relevant fluctuations would be between $|N=4\rangle$ and $|N=2,n_1n_2=0\rangle$, but this low-spin doublon state is suppressed for $J>0$.) The photo-doped single-orbital model is interesting because the gap-size in this model is not strongly affected by the insertion of doublons and holons, which allows to study the spin singlet pairing susceptibility in a system with intermediate $U$. A previous study\cite{Rosch2008} has already predicted the appearance of a metastable condensate in a large-$U$ Hubbard system consisting of doublons moving in a background of empty sites. A relevant open questions is the pairing tendency in a photo-doped state where doublons and holons coexist with singly occupied sites. While the cooling by local spin excitations is absent in this model, an enhanced pairing susceptibility can be expected in the presence of a bosonic heat bath.

\acknowledgements
The calculations have been performed on the Beo04 computer cluster at the University of Fribourg. We acknowledge financial support from ERC Consolidator Grant No.~724103, ERC Starting Grant No.~716648, and from the Swiss National Science Foundation through NCCR Marvel. 
 
\appendix 
 
\section{Derivation of the DMFT self-consistency equation}
\label{appendix}

The DMFT self-consistency equation (\ref{self_main}) can be derived using the cumulant expansion.\cite{Strand2015} We decompose the lattice action into the contribution of site 0, a cavity action which describes the lattice without site 0, and a hopping term which connects the two: $S_\text{latt}=S_0+S^{(0)}+\Delta S$ with $\Delta S= \int dt \sum_j (\psi^\dagger_0 V^*_{j} \psi_j  + \psi^\dagger_j V_{j} \psi_0)\equiv \int dt \sum_j(\Delta H_{0j} + \Delta H_{j0})\equiv \int dt \Delta H $. After cumulant expansion of $\langle e^{-i(S_0+\Delta S)}\rangle_{S^{(0)}}$ and re-exponentiation we obtain the expression for the effective action
\begin{align}
-iS=&-iS_0+\sum_{n=1}^\infty \frac{1}{n!}\int d t_1 \ldots d t_n \nonumber\\
&\hspace{9.5mm}\times \langle (-i\Delta H)(t_1) \ldots (-i\Delta H)(t_n)\rangle_{S^{(0)}},
\end{align}
which in DMFT can be truncated at order $n=2$. Thus, the term to evaluate is 
\begin{equation}
-\frac{i}{2}\sum_{ij}\langle 
(\Delta H_{0i}(t)+\Delta H_{i0}(t)) (\Delta H_{0j}(t')+\Delta H_{j0}(t'))\rangle_{S^{(0)}}.
\label{secondorder}
\end{equation}
The contributions $\langle \Delta H_{0i}(t)\Delta H_{0j}(t') \rangle_{S^{(0)}}$ and $\langle \Delta H_{i0}(t)\Delta H_{j0}(t') \rangle_{S^{(0)}}$ vanish if we consider neither intra-orbital pairing, nor conventional inter-orbital hybridizations (excitonic order). $\langle \Delta H_{i0}(t)\Delta H_{0j}(t') \rangle_{S^{(0)}}$ gives the same contribution as $\langle \Delta H_{0i}(t)\Delta H_{j0}(t') \rangle_{S^{(0)}}$ (after exchanging $t\leftrightarrow t'$ and $i\leftrightarrow j$). Equation (\ref{secondorder}) thus evaluates to 
\begin{align}
&-i\sum_{ij}\langle \Delta H_{0i}(t)\Delta H_{j0}(t') \rangle_{S^{(0)}}\nonumber\\
&=(-i)\sum_{ij}\langle \psi^\dagger_0(t) V^*_{i}(t) \psi_i (t) \psi^\dagger_j(t') V_{j}(t') \psi_0 (t')\rangle_{S^{(0)}}\nonumber\\
&=(-i)\Big[\sum_{ij,\sigma} c^\dagger _{10\sigma}(t) v^1_{0i}(t) \langle c_{1i\sigma}(t) c^\dagger _{1j\sigma}(t')\rangle_{S^{(0)}} v^1_{j0}(t')c_{10\sigma}(t')\nonumber\\
&\hspace{4mm}+\sum_{ij,\sigma} c_{20\sigma}(t) v^2_{i0}(t) \langle c_{2i\sigma}^\dagger (t) c_{2j\sigma}(t')\rangle_{S^{(0)}} v^2_{0j}(t')c^\dagger _{20\sigma}(t')\nonumber\\
&\hspace{4mm}-\sum_{ij,\sigma\sigma'} c_{20\sigma}(t) v^2_{i0}(t) \langle c^\dagger _{2i\sigma}(t) c^\dagger _{1j\sigma'}(t')\rangle_{S^{(0)}} v^1_{j0}(t')c_{10\sigma'}(t')\nonumber\\
&\hspace{4mm}-\sum_{ij,\sigma\sigma'} c^\dagger _{10\sigma}(t) v^1_{0i}(t) \langle c_{1i\sigma}(t) c_{2j\sigma'}(t')\rangle_{S^{(0)}} v^2_{0j}(t')c^\dagger_{20\sigma'}(t')\Big].
\end{align}
(Note the minus signs on the anomalous terms.) 
On a Bethe lattice, we have the constraint $i=j$. By defining the cavity Green's function $G^{ab (0)}_{ij,\alpha \sigma \beta\sigma'}=-i\langle \mathcal{T}a_{i,\alpha\sigma}(t) b_{j,\beta \sigma'}(t')\rangle_{S^{(0)}}$ we thus see that the second order contribution to the effective action is given by the hybridization term $\int_{\mathcal C} dt dt' \psi^\dagger(t)\Delta(t,t')\psi(t')$ with the hybridization function defined in Eq.~(\ref{self_main}).  
In DMFT, we furthermore replace the cavity Green's functions $G^{ab(0)}_{jj,\alpha\sigma\beta\sigma'}$ by the lattice Green's function 
$G^{ab}_{jj,\alpha\sigma\beta\sigma'}\equiv G^{ab}_{\alpha\sigma\beta\sigma'}$, 
so that the self-consistency equation becomes
\begin{equation}
\Delta(t,t') = \sum_{j=1}^z V_j^*(t) G(t,t') V_j(t'),
\label{self}
\end{equation}
with $V_j(t)=\text{diag}(v^1_{j0}(t), -v^2_{0j}(t), v^1_{j0}(t), -v^2_{0j}(t))$.
 
Following Ref.~\onlinecite{Werner2017} we can also derive a self-consistency condition for a ``Bethe lattice with electric field" \cite{Werner2017} by considering a chain ($z=2$) and the complex hoppings $v^\alpha_{j0}(t) = (v^\alpha_{0j})^*(t)=v^\alpha(t) e^{i\phi_\alpha(t)}$, with $v^\alpha(t)$ real and $\phi_\alpha(t)=\int_0^t dt' A_\alpha(t')$ the integral of the vector potential along the chain:
\begin{equation}
\Delta(t,t') = W^*(t) G(t,t') W(t') + W(t) G(t,t') W^*(t'),\nonumber
\end{equation}

\noindent
where $W(t)=\text{diag}(v^1(t)e^{i\phi_1(t)}, -v^2(t)e^{-i\phi_2(t)}, v^1(t)e^{i\phi_1(t)},$ $ -v^2(t)e^{-i\phi_2(t)})$. In order to recover the usual $z=\infty$ Bethe lattice selfconsistency in the case without field, we have to set $v^\alpha = v/\sqrt{2}$. 
 


\begin{thebibliography}{99}
\bibitem{footnote_component}{We write the formulas for the ``$1\!\uparrow$" component, but the results are identical for the other spin and orbital components.}
\end{thebibliography}
\end{document}